\documentclass[pra,twocolumn,floatfix,superscriptaddress,longbibliography,notitlepage]{revtex4-2}
\usepackage{amssymb,amsmath,amsthm,color,graphicx,times,graphicx}
\usepackage{caption}
\usepackage{ragged2e}
\DeclareCaptionJustification{justified}{\justifying}
\usepackage{hyperref}
\usepackage{braket}
\usepackage{graphicx}% Include figure files
\usepackage{dcolumn}% Align table columns on decimal point
\usepackage{appendix}
\usepackage{subcaption}
\usepackage{xcolor}
\usepackage{bm}% bold math
\DeclareMathOperator{\Tr}{Tr}
\providecommand{\openone}{\leavevmode\hbox{\small1\kern-4.3pt\normalsize1}}
\renewcommand\Re{\mathrm{Re}}

\renewcommand{\Re}{\mathrm{Re}}
\renewcommand{\Im}{\mathrm{Im}}

 \usepackage{orcidlink}

\theoremstyle{plain}

\theoremstyle{definition}

\begin{document}
\title{Multiparticle Quantum Heat Engine: Exploring the Impact of Criticality on Efficiency}

\author{Anass Hminat \orcidlink{0009-0007-3677-3952}}\affiliation{LPHE-Modeling and Simulation, Faculty of Sciences, Mohammed V University in Rabat, Rabat, Morocco.}
\author{Abdallah Slaoui \orcidlink{0000-0002-5284-3240}}\email{Corresponding author: abdallah.slaoui@fsr.um5.ac.ma}\affiliation{LPHE-Modeling and Simulation, Faculty of Sciences, Mohammed V University in Rabat, Rabat, Morocco.}\affiliation{Centre of Physics and Mathematics, CPM, Faculty of Sciences, Mohammed V University in Rabat, Rabat, Morocco.}
\author{Brahim Amghar \orcidlink{0009-0008-8209-9644}}\affiliation{Laboratory LPNAMME, Laser Physics Group, Department of Physics, Faculty of Sciences, Chouaïb Doukkali University, El Jadida, Morocco.}\affiliation{Centre of Physics and Mathematics, CPM, Faculty of Sciences, Mohammed V University in Rabat, Rabat, Morocco.}
\author{Rachid Ahl Laamara \orcidlink{0000-0002-8254-9085}}\affiliation{LPHE-Modeling and Simulation, Faculty of Sciences, Mohammed V University in Rabat, Rabat, Morocco.}\affiliation{Centre of Physics and Mathematics, CPM, Faculty of Sciences, Mohammed V University in Rabat, Rabat, Morocco.}

\begin{abstract}
Quantum many-body systems present substantial technical challenges from both analytical and numerical perspectives.  Despite these difficulties, some progress has been made, including studies of interacting atomic gases and interacting quantum spins.  Furthermore, the potential for criticality to enhance engine performance has been demonstrated, suggesting a promising direction for future investigation. Here, we explore the performance of a quantum Otto cycle using a long-range Ising chain as the working substance.  We consider an idealized cycle consisting of two adiabatic transformations and two perfect thermalizations, eliminating dissipation. Analyzing both engine and refrigerator modes, we investigate the influence of particle number, varied from $10$ to $100$, on efficiencies and behavior near the critical point of the phase transition, which we characterize using a scaling factor. We also examine how internal factors—specifically, the power-law exponent, the number of particles, and the hot and cold reservoir temperatures—affect the system's operation in different modes.  Our results reveal that these factors have a different impact compared to their classical counterparts. 
\par
\vspace{0.25cm}
%\textbf{Keywords:}Thermodynamic engine, Quantum heat engine, work scaling factor, phase transition, quantum criticality, disorder.
%\pacs{04.79.Ta, 04.78.Uz, 04.57.Sn, 42.53.-p, 03.65.Ud}
\end{abstract}
\date{\today}

\maketitle
\section{Introduction}
Thermodynamics is a remarkable theory in physics, primarily aimed at studying heat \cite{Winterbone2015}. It has been instrumental in the design of internal combustion engines and even in explaining the behavior of black holes. Technological progress increasingly requires the miniaturization of components \cite{Deffner2019}, pushing them to a regime where quantum effects are no longer negligible. The necessity of understanding the underlying mechanisms of heat conversion into useful work has led researchers to formulate the three laws of thermodynamics. More recently, with the rise of quantum technologies and the miniaturization of devices that exchange heat and work on the nanoscale, it has become increasingly relevant to investigate these mechanisms within the framework of quantum mechanics \cite{Strasberg}. However, while classical thermodynamics is a well-established theory, its extension to the quantum domain presents conceptual challenges and remains an active area of research. Pioneering efforts to generalize the concepts of heat and work for quantum systems date back to the 1980s \cite{Perez}. Only recently has a renewed interest in quantum thermal machines spurred prolific scientific advancements, starting with the groundbreaking proposal of the maser \cite{Scovil1959} as the first example of a quantum machine.\par

A quantum thermal machine is broadly defined as a device made of quantum systems capable of performing work by undergoing a suitable thermodynamic cycle. Several studies have proposed potential implementations of few-body thermal machines \cite{Latifah2013} based on the Carnot \cite{Bender2000, Bender2002}, Otto \cite{Solfanelli2020, Piccitto2022}, Stirling \cite{Yin2017, Yin2018}, Brayton \cite{Setyo2018a}, and Diesel cycles \cite{Setyo2018}. Some of these realizations have been achieved using various quantum systems, including infinite potential wells \cite{Purwanto2016}, nuclear magnetic resonance (NMR) \cite{Camati}, ultra-cold atoms \cite{Bouton2021}. However, the role played by many-body interactions in the thermodynamic performance of quantum machines \cite{Arezzo2024,MakouriSlaoui2025,MakouriSlaoui2024} is not yet fully understood. Determining whether quantum machines with many-body interactions can outperform classical ones, and under what conditions this occurs, is currently at the forefront of debates in quantum thermodynamics. Long-range interacting systems \cite{Ranabhat2023,Amghar2023} offer promising prospects for quantum technological applications due to their resilience to external disturbances \cite{Lerose2019}. This stability enables control over the impact of dynamically generated excitation, thereby mitigating their detrimental effects. Specifically, the dynamical phase transition \cite{Piccitto2019,Piccitto2019a} is clear when $\alpha < 1$. But when $\alpha > 1$, the critical point grows into a chaotic crossover region where the dynamics and asymptotic state are very dependent on the system's parameters.A promising direction for exploring quantum advantage lies in using many-body quantum systems as the working substance in thermodynamic engines \cite{MurherjeeV,CangemiLM}. Although this approach is hindered by technical challenges, both analytically and numerically, some preliminary results have been obtained \cite{JamilloM,WatsonRS,Halpernny}. It has been shown that quantum criticality can play a crucial role in enhancing engine performance \cite{Carollof,Zhang2007,Munoz2014}. However, the exact role that many-body interactions play in the thermodynamic efficiency of quantum engines remains poorly understood.
\par
The study of topological phases in systems with long-range (LR) interactions represents a dynamic research frontier, bolstered by recent experimental proposals. Notably,  chains with non-local hopping and pairing are promising candidates for realization 
\cite{Xu2022} in solid-state systems,an
example of the rigidity of long-range interacting platforms against external drivings, is the possibility for
such systems to host Floquet time crystal phases \cite{Russomanno2017,Surace2019,Pizzi2021}. Theoretical insights affirm that symmetry-protected topological order can persist under LR interactions \cite{Defenu2019} in specific systems. However, the generalizability of these findings to broader symmetry-protected topological phases \cite{Pizzi2021,Defenu2019,Halimeh2020,Acevedo2014,Hwang2015,Defenu2018} with LR interactions remains an open question, necessitating further investigation. These systems give the possibility to
study a rich variety of phenomena which contrast the
non-equilibrium dynamics of systems with long-range interactions to that of short-range ones, such as the violation of Lieb-Robinson bounds and anomalous propagation of information \cite{Richerme2014,Jurcevic2014,Koffel2012,Buyskikh2016,Tarif2025}.

\par
In this paper, we contribute to this burgeoning field by presenting an in-depth study of the quantum Otto engine based on the Ising model (see Fig.\ref{fig1}), focusing particularly on the potential influence of quantum criticality on the engine's thermodynamic performance.Our analysis also reveals that the quantum Otto engine based on the Ising model can function either as a heat engine—extracting work by absorbing heat from the cold reservoir—or as a refrigerator—using input work to transfer heat from the cold to the hot reservoir—depending on the specific parameters governing the thermodynamic cycle. While the operational regime is indeed sensitive to these parameters, one can always identify a suitable parameter range in which each mode of operation is achieved with high reliability. This study rigorously explores a quantum thermodynamic system governed by the Otto cycle, highlighting its quantum features distinct from classical counterparts \cite{Campisi2016}. We first outline the system and cycle, examining the impact of parameters such as system size \( N \), initial field \( h_i \), interaction exponent \( \alpha \), and inverse temperatures \( \beta_c \), \( \beta_h \), on operational modes, including transitions between refrigerator and engine regimes. We then assess the quantum refrigerator and thermal engine, analyzing how \( N \) and thermal correlations influence refrigeration capacity \( Q_c / N \), work output \( W / N \), and efficiencies \( \eta_R \), \( \eta \). With fixed adiabatic compression time, we investigate idealized machines with infinitely slow phases, quantifying work per spin, efficiency, and stability via the scaling factor \( \Pi_R / N \) or \( \Pi / N \) \cite{Campisi2016}. Additionally, we probe correlations for \( \alpha \in [0.01, 1.3] \) at \( N = 100 \), evaluating efficiency variations with \( Q_c / N \), \( W / N \), and \( h_i \), and their distribution across chain sizes \cite{Piccitto2019,Piccitto2019a}. Lastly, we examine the scaling factor proportionality for \( N \in [10, 100] \), elucidating critical behaviors near phase transitions.
\begin{figure}[h]
    \centering
\includegraphics[width=1.0\linewidth]{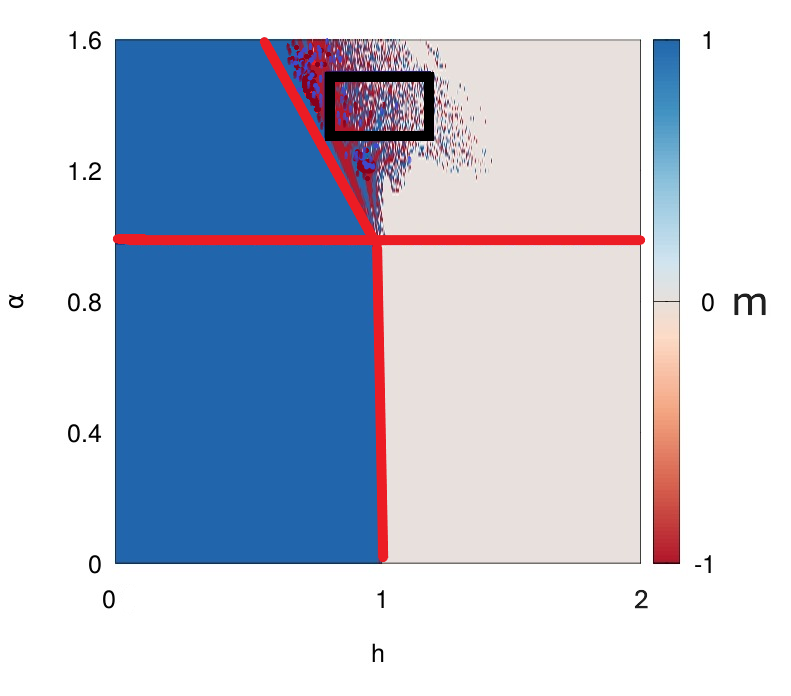} 
    \caption{Dynamical phase diagram for the long-range Ising model ($J = 1$), obtained via CMFT with $\ell = 4$, following a quench of the transverse field from $h_{\text{in}} = 0$ to $h$. The plot shows the time-averaged longitudinal magnetization (up to a time $T = 1000$) as a function of $\alpha$ and $h$. Note that, as soon as $\alpha > 1$ (indicated by the dashed line), the dynamical phase transition evolves into a critical region, where the asymptotic magnetization becomes highly sensitive to the system parameters. The blue square highlights.}
    \label{1}
\end{figure}
\section{quantum otto cycle}
In this section, we will describe the system and outline its phase transition. We will then examine the thermodynamic cycle and the associated physical quantities, such as the heat exchanged and the work done. Afterward, we will explore how the interaction range affects the efficiency of the cycle. The analysis will focus on how both short-range and long-range interactions influence the system’s performance, particularly near the critical point.
\subsection{Many-body quantum Ising model}

We study the one-dimensional transverse-field Ising model with long-range (LRIM) power-law decaying interactions. We consider an open boundary condition, which is described by the following Hamiltonian
\begin{equation}
H =- \frac{1}{K(\alpha)} \sum_{i<j}^{N} \frac{J}{|i - j|^\alpha} \hat{s}_i^x \hat{s}_j^x - h \sum_{i=1}^{N} \hat{s}_i^{z},
\end{equation}

where $s_{i}^\mu$ are spin matrices acting on site $i$ and $\alpha$ is the power-law exponent. The Kac normalization constant, defined as
\begin{equation}
K(\alpha) = \frac{1}{N - 1} \sum_{i < j} \frac{1}{|i - j|^\alpha},
\end{equation}
ensures that the energy density remains intensive for $\alpha \leq 1$. For $\alpha = \infty$, this model reduces to the standard transverse field Ising model (TFIM), which can be exactly solved using the Jordan-Wigner transformation \cite{Lieb1961,Mbeng2024} . The long range exhibits a quantum phase transition from the ferromagnetic to the paramagnetic phase at $h_c=J$. This transition persists as $\alpha$ decreases \cite{Koffel2012}, with the transition point shifting towards higher magnetic field values. Thus\par

\begin{equation}
H = -\sum_{j=1}^{N} \sum_{r=1}^{N/2-1} \left[ t_r c^\dagger_{j+r} c_j + \Delta_r c^\dagger_{j+r} c^\dagger_j + \text{h.c.} \right] - h \sum_{j=1}^{N} \left[ 1 - 2 c^\dagger_j c_j \right], \label{eq:hamiltonian}
\end{equation}
where $c^\dagger_j$ and $c_j$ are creation and annihilation operators for fermions at site $j$, while $t_r$ and $\Delta_r$ are the hopping and pairing amplitudes, respectively. We choose their dependence on the intersite distance $r$ according to the power laws
\begin{equation}
t_r = \frac{1}{N_{\alpha_1}} \frac{J}{r^{\alpha_1}}, \quad \Delta_r = \frac{1}{N_{\alpha_2}} \frac{J}{r^{\alpha_2}}, \label{eq:power_laws}
\end{equation}
\begin{figure}[h]
    \centering
    \includegraphics[width=1.0\linewidth]{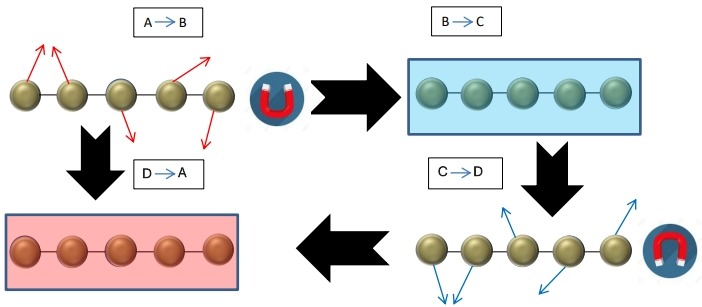} 
    \caption{The schematic of the Ising Long-range quantum Otto engine consists of two adiabatic phases and two thermalization phases. From (A → B), the system undergoes an adiabatic process where the transverse field increases from \(h_i\) to \(h_f\). In (B → C), the system thermalizes at temperature \(T_h\). Then, from (C → D), it is adiabatically brought back to \(h_i\), followed by thermalization at temperature \(T_c\) in (D → A), completing the cycle.}
    \label{fig1}
\end{figure}

with the hopping exponent $\alpha_1 > 0$, the pairing exponent $\alpha_2 > 0$, and $N_\alpha = \sum_{r=1}^{N/2} r^{-\alpha}$ the Kac scaling factor \cite{Piccitto2019a,Piccitto2019}, which guarantees extensivity of the energy in the case $\alpha_i < 1$, with $i = 1, 2$. Hereafter, we set $J = \Delta = 1$ as the energy scale and work in units of $\hbar = k_B = 1$. This model, and with same range for pairing and hopping $\alpha_1 = \alpha_2 = \alpha$ \cite{Solfanelli2020}.

Due to the translational invariant nature of the couplings, it is useful to write the Hamiltonian in terms of the momentum-space operators
\begin{equation}
\tilde{c}_k = e^{-i \frac{\pi}{4}} \frac{1}{\sqrt{N}} \sum_{j=1}^N e^{ikj} c_j, \label{eq:momentum_operators}
\end{equation}
where $k = 2\pi n / N$, with $n = -N/2 + 1, \dots, N/2$ (in the following we will drop the $\tilde{}$ on the $c_k$ unless it is ambiguous). Then we obtain
\begin{equation}
H = \sum_k \left[ (h - t_k) \left( c^\dagger_k c_k - c_{-k} c^\dagger_{-k} \right) + \Delta_k \left( c^\dagger_k c^\dagger_{-k} + c_{-k} c_k \right) \right], \label{eq:hamiltonian_momentum}
\end{equation}
where $t_k$ and $\Delta_k$ are the Fourier transforms of the hopping and pairing amplitudes, respectively, which in the thermodynamic limit may be written as
\begin{align}
t_k &= \frac{\Re \left[ \text{Li}_{\alpha_1} \left( e^{ik} \right) \right]}{\zeta(\alpha_1)}, \label{eq:tk} \\
\Delta_k &= \frac{\Im \left[ \text{Li}_{\alpha_1} \left( e^{ik} \right) \right]}{\zeta(\alpha_1)}, \label{eq:delta_k}
\end{align}
where $\text{Li}_\alpha(z)$ denotes the polylogarithm and $\zeta(\alpha)$ is the Riemann zeta function. We notice that the Hamiltonian in the Fourier space can be decomposed into the sum of single mode Hamiltonians, introducing $\Psi_k = (c_k, c^\dagger_{-k})^T$
\begin{equation}
H = \sum_k \Psi^\dagger_k H_k \Psi_k, \label{eq:hamiltonian_decomposed}
\end{equation}
\begin{equation}
H_k = (h - t_k) \sigma^z_k + \Delta_k \sigma^x_k, \label{eq:mode_hamiltonian}
\end{equation}
where $\sigma^{(a)}_k$, $a = x, y, z$ are the Pauli sigma operators. Let us notice how the $k$th term of the Hamiltonian acts on a different sector of the total Hilbert space, namely the two-dimensional subspace spanned by the states $\ket{0_k, 0_{-k}}$, $\ket{1_k, 1_{-k}} = c^\dagger_k c^\dagger_{-k} \ket{0_k, 0_{-k}}$. Then, the Hamiltonian is diagonalized via a Bogoliubov transformation, in terms of the fermionic quasiparticle operators $\gamma_k = u_k c_k + v^*_{-k} c^\dagger_{-k}$, with Bogoliubov coefficients
\begin{equation}
u_k = \cos \frac{\theta_k}{2}, \quad v_k = \sin \frac{\theta_k}{2}, \label{eq:bogoliubov_coeffs}
\end{equation}
where $\theta_k = \arctan \left[ \Delta_k / (h - t_k) \right]$, to obtain
\begin{equation}
H = \sum_k \omega_k(h) \left( \gamma^\dagger_k \gamma_k - \frac{1}{2} \right), \label{eq:diagonalized_hamiltonian}
\end{equation}
with the spectrum
\begin{equation}
\omega_k(h) = 2 \sqrt{(h - t_k)^2 + \Delta_k^2}. \label{eq:spectrum}
\end{equation}

At the opposite extreme, $\alpha=0$ , represents a fully connected regime, which can be analytically treated for its static and dynamic properties. In this case, the model shows long-range ferromagnetic order at low finite temperatures. This configuration is particularly interesting for $\alpha < 2$, revealing exotic phenomena such as prethermalization  \cite{Ranabhat2023,Ranabhat2024}, dynamical phase transitions \cite{Piccitto2019a,Piccitto2019}, and dynamical confinement \cite{Liu2019tt,Scopa2022}.\par

This model is strongly influenced by the parameter $\alpha$. For instance, for $\alpha < 2$, long-range ferromagnetic order is observed at finite temperatures \cite{Gonzalez2021}. The critical magnetic field is estimated using the Cluster approach and Mean-Field Theory (CMFT) \cite{jin2016,ElMakouri2023,yamamoto2009} to explore the phase diagram of an Ising spin chain with long-range interactions. This method captures short-range correlations while treating interactions between clusters at a mean-field level, revealing the emergence of a chaotic region for specific values of the parameter \(\alpha\). In our study, we avoid mean-field or cluster decomposition approximations and spin interaction is uniform. Nevertheless, we observe significant concordance with the results obtained via CMFT, especially concerning static phase transitions, it is possible to extract a qualitative behavior of the quantum critical point as a function of \(\alpha\). Interpolating the data, we find
\[
h_c(\alpha) \sim 
\begin{cases} 
1 & \text{if } \alpha \leq 1, \\
0.35(3.2 - \alpha) & \text{if } 1 < \alpha < 2.
\end{cases}
\]
The nonequilibrium dynamics following a quench may be probed through the time-averaged longitudinal magnetization, defined as \( m_z = \lim_{T \to \infty} \frac{1}{T} \int_0^T \langle \sigma_z(t) \rangle \, dt \), serving as the order parameter. Initiating from a fully polarized state, the system undergoes a dynamical quantum phase transition at a critical field \( h_c = J \). In the thermodynamic limit Fig.\ref{1}, for \( 0 < \alpha < 1 \), the long-range Ising chain exhibits behavior analogous to the Lipkin-Meshkov-Glick model. However, as the power-law exponent surpasses the critical threshold \( \alpha = 1 \), fluctuations become increasingly pivotal, profoundly shaping the collective dynamics of the system.

\par
Since our method is essentially a generalized mean field theory, we will use it when the latter describes the equilibrium transition (\(\alpha < 5/3\)). Despite this, it must be stressed that CMFT is exact for large \(\ell\). We will use a numerical diagonalization of the Hamiltonian in the disordered case (Consult appendix ~\ref{appendixA} and~\ref{appendixB}for more details) .similar to  \cite{Piccitto2022}, where we will assume $J_{ij}=\frac{1}{K(\alpha)}\frac{J}{|i - j|}$, to compute $\omega_k(h)$ for more details consult Otto \cite{Mbeng2024}. In conclusion, the transverse field Ising model with power-law decaying interactions offers a rich theoretical framework for exploring quantum phase transitions and complex dynamics while being directly applicable to concrete experimental setups.

\subsection{Cycle description}
 We will work on an ideal otto cycle 
illustrated in  Fig.\ref{fig1} composed of two infinitely slow adiabatic phases and two thermalization phases without any dissipation. After that, we will examine the influence of the thermal correlations of the $N$ particles on the efficiency of the cycle. For the dynamic and ideal case, we have:\par

\begin{itemize}
    \item[(a)]{\bf \( A \to B \):  Adiabatic field increase;} \\ The transverse field is linearly ramped up from \( h_i \) to \( h_f \) over time, while the working substance remains isolated from the thermal baths.
    \[
h(t) = h_i + v t, \quad t \in \left[0, \frac{h_f - h_i}{v} \right]
\]

    \item[(b)]{\bf \(B \to C\):  Thermalization with the cold bath;}\\ The Hamiltonian \( \hat{H}_{\text{sys}}(t_f) \) is held constant, and the system is coupled to the cold bath until the working substance reaches the thermal state of \( \hat{H}_{\text{sys}}(t_f) \) at temperature \( T_c \).
    \item[(c)]{\bf \( C \to D \): Adiabatic field decrease;} \\ The transverse field is linearly reduced from \( h_f \) back to \( h_i \) at the same rate as in step (a), with the working substance decoupled from the baths.
    \[
h(t) = h_f - v t, \quad t \in \left[0, \frac{h_f - h_i}{v} \right]
\]

    \item[(d)] {\bf\(  D \to A \): Thermalization with the hot bath;} \\
  The Hamiltonian \( \hat{H}_{\text{sys}}(t_i) \) is maintained, and the system is connected to the hot bath until the working substance returns to its initial thermal state of \( \hat{H}_{\text{sys}}(t_i) \) at temperature \( T_h \).
\end{itemize}

\begin{figure}[h]
    \centering
    \includegraphics[width=0.6\linewidth]{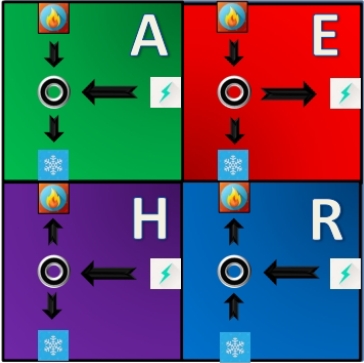} 
    \caption{The color code represents the following modes: green for accelerator, purple for heater, red for heat engine, and blue for refrigerator.}
    \label{fig2}
 \end{figure}quantities based on the Appendix A \cite{Munoz2012,Chattopadhyay2019}. The efficiency of the cycle can thus be expressed as:
\begin{equation}
\eta = \frac{W}{Q_h}.
\end{equation}

In what follows, we use \( Q_c(h) \) to represent the heat exchanged during thermalization with the cold (or hot) bath, adopting the convention that \( Q_c(h) > 0 \) when the system absorbs heat from the reservoir. If \( \rho_\alpha \) is the density matrix of the system at points \( \alpha = A, B, C, D \), then:
\begin{align}
    Q_c &= \langle \hat{H}(t_f) \rangle_{\rho_C} - \langle \hat{H}(t_f) \rangle_{\rho_B}, \label{eq:Qc} \\
    Q_h &= \langle \hat{H}(t_i) \rangle_{\rho_A} - \langle \hat{H}(t_i) \rangle_{\rho_D}. \label{eq:Qh}
\end{align}
It describes a one-dimensional long-range Ising system of \( N \) quantum spins interacting with a ferromagnetic coupling strength \( J > 0 \), in the presence of magnetic field \( h \). Hereafter, we set    \( J = 1.0 \)   as the energy scale and work in units where $\hbar = k_{B} = 1$.The long-range interaction between spins is dictated by the strength parameter \( \alpha \) (we set \(h_{i} < h_{f} \) and \( h_{f} - h_{i} = 0.5 \)).
When analyzing the case of an infinitely slow cycle, i.e., the limit $T \to \infty$, the regime is typically referred to as adiabatic. In this regime, the unitary evolution is slow enough to satisfy the adiabatic theorem, preventing transitions between the instantaneous eigenstates of the Hamiltonian. The nonlocal master equation chosen to model the
system–environment interaction \cite{DABBRUZZO}. Let us consider, for a while, the more general configuration of \( N_B \)
independent thermal reservoirs at temperature \( T_n \), with \( n \in \{1, \dots, N_B\} \) indices labeling the bath.
The Hamiltonian describing this setup reads

\[
\hat{H}_{\text{env}} =
\sum_{n=1}^{N_B} \int dk \, \varepsilon_n(k) \hat{c}_n^\dagger(k) \hat{c}_n(k),
\]

with \( \varepsilon_n(k) \geq 0 \) and \( \hat{c}_n(k) \), \( \hat{c}_n^\dagger(k) \) being fermionic annihilation and creation operators. The \( N_B \) baths are independent,
therefore the corresponding reduced density operator of the full environment assumes the factorized form

\[
\rho_{\text{env}} = \bigotimes_{n=1}^{N_B} \rho_{\text{bath}}^{(n)},
\]

where \( \rho_{\text{bath}}^{(n)} \) is the thermal density matrix describing the \( n \)th fermionic bath at temperature \( T_n \).

\begin{figure*}[t] 
    \centering

    \begin{subfigure}[t]{0.33\textwidth} % Width allocation for each subfigure
 
        \centering
   
       \raisebox{-0.2cm}{\includegraphics[width=6.7cm]{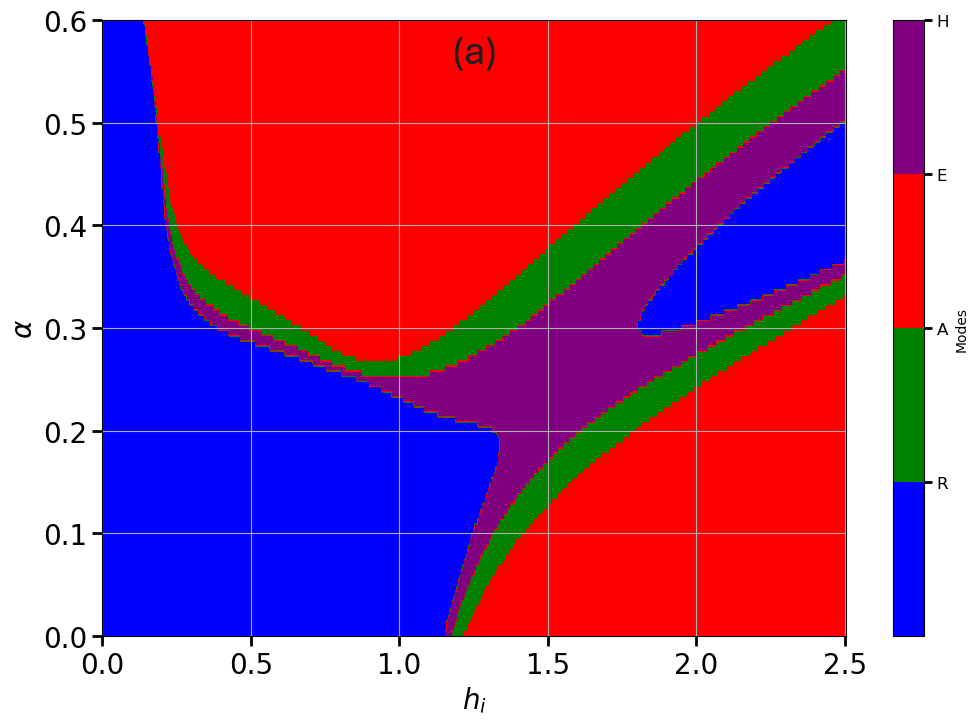}} % Manually set width to 10cm
    \end{subfigure}
    \hfill
    \begin{subfigure}[t]{0.33\textwidth} % Width allocation for each subfigure
        \centering
        \raisebox{-0.15cm}{\includegraphics[width=6.74cm]{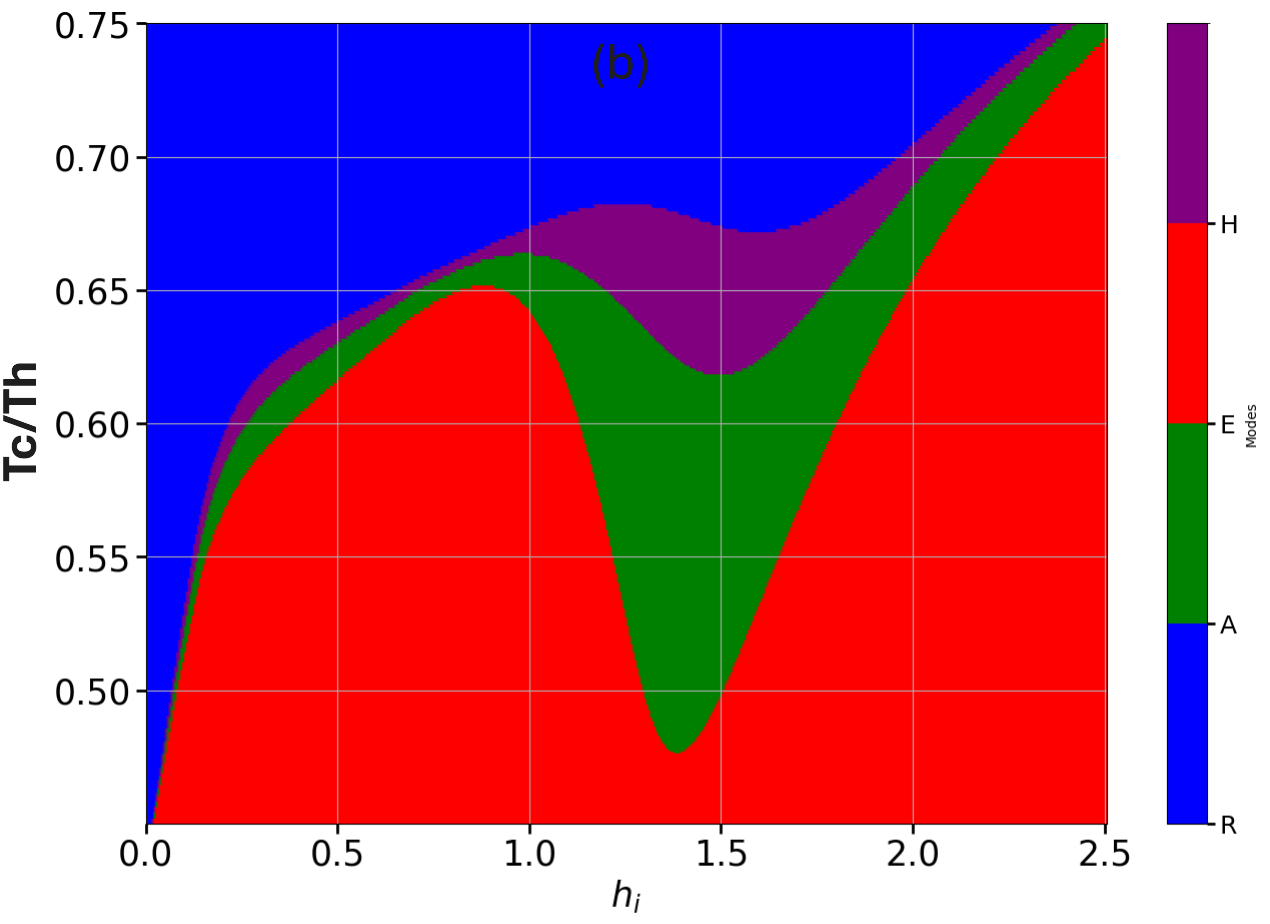} }% Manually set width to 10cm

    \end{subfigure}
     \hfill
    % Second row with single centered image
    \begin{subfigure}[t]{0.33\textwidth} % Width allocation for centered subfigure
        \centering
         \raisebox{-0.22cm}{\includegraphics[width=6.7cm]{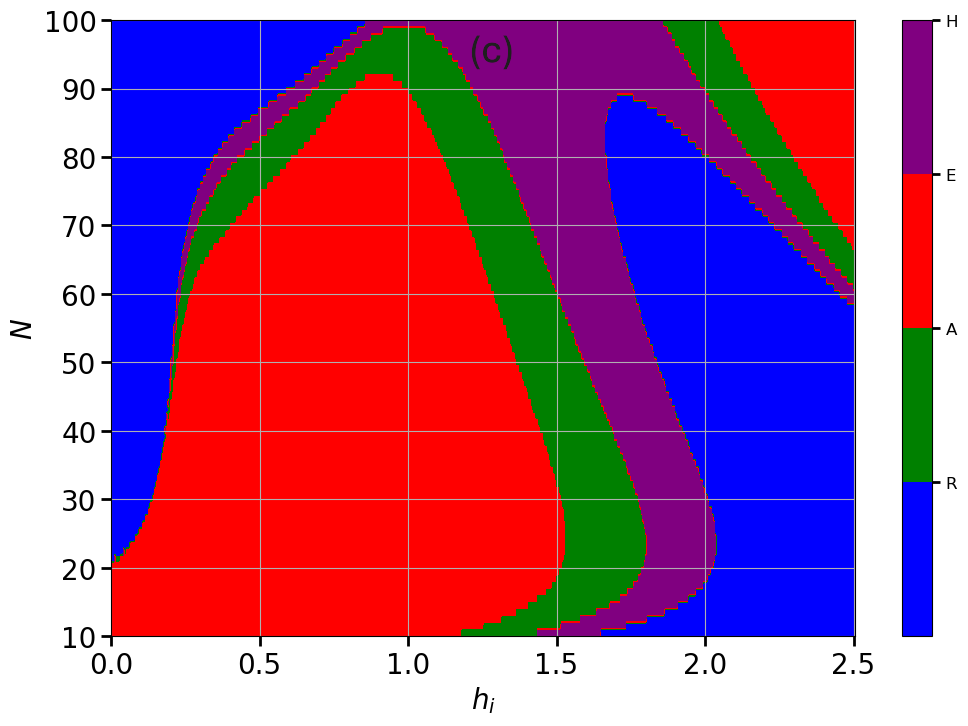}} % Set width for centered image
    \end{subfigure}

    \caption{Operating regimes of the quantum Otto engine:  
(a) in the $\alpha$–$h_i$ plane, with $T_h = 0.57$ and $T_c = 0.38$, and $\alpha$ ranging from 0 to 0.6;  
(b) in the $T_c/T_h$–$h_i$ plane, with $T_h = 1.0$ and $T_c = 0.75$, and $\gamma$ ranging from $-1$ to $1$;  
(c) in the $N$–$h_i$ plane, with the temperature ratio $T_c/T_h$ varying from 0.5 to 1 and $\Gamma$ fixed at 0;  
 where $\delta h = 0.5$ is kept fixed.The color code represents the following modes: accelerator (green), heater (purple), heat engine (red), and refrigerator (blue).}
\label{Fig3}
\end{figure*}

Tracing out all the environmental degrees of freedom and imposing the Born–Markov approximation
for the baths, it is possible to derive a microscopic Lindblad master equation , in the energy eigenbasis,
for the reduced density matrix of the system described by the Hamiltonian :

\[
\frac{d}{dt} \rho_{\text{sys}}(t) = -i [\hat{H}_{\text{sys}}, \rho_{\text{sys}}] + D[\rho_{\text{sys}}],
\]

\begin{equation}
\begin{aligned}
D[\rho_{\text{sys}}] =
\sum_{n,k} \gamma_{nk} & \bigg[ (1-f_n(\omega_k)) 
\left( \frac{1}{2} \hat{b}_k \rho_{\text{sys}} \hat{b}_k^\dagger 
- \frac{1}{2} \{\hat{b}_k^\dagger \hat{b}_k, \rho_{\text{sys}}\} \right)  \\
& + f_n(\omega_k) \left( \frac{1}{2} \hat{b}_k^\dagger \rho_{\text{sys}} \hat{b}_k 
- \frac{1}{2} \{\hat{b}_k \hat{b}_k^\dagger, \rho_{\text{sys}}\} \right) \bigg].
\end{aligned}
\end{equation}

where the \( \{\hat{b}_k, \hat{b}_k^\dagger\} \) jump operators are the fermionic Bogoliubov quasiparticles which diagonalize the model
, these operators are local in the energy eigenbasis and thus nonlocal in the
sites, giving rise to a global master equation  \cite{DABBRUZZO}. Moreover,

\[
f_n(\omega_k) = \frac{1}{1 + e^{\omega_k/T_n}},
\]

Under the assumption of no degeneracies in the spectrum (as turns out to be the case for the Ising
chain with open boundary conditions), it can be used to obtain an analytic expression for the
time evolution of the correlation functions. In particular, defining 

\[
\tilde{f}_k = \frac{\sum_n \gamma_{nk} f_n(\omega_k)}{\sum_n \gamma_{nk}},
\]

we have

\[
\langle \hat{b}_k^\dagger \hat{b}_k \rangle_t = \tilde{f}_k \left( 1 - e^{-2 \sum_n \gamma_{nk} t} \right) + \langle \hat{b}_k^\dagger \hat{b}_k \rangle_0 e^{-2 \sum_n \gamma_{nk} t}.
\]

Then, all the thermodynamic quantities can be easily computed using equation (20) or, equivalently, directly computing the partition function \( Z \). In particular, the internal energy reads:

\begin{equation}
 \langle E_i  \rangle = \sum_k \omega_{k,i} \left( \langle b_k^\dagger b_k \rangle - \frac{1}{2} \right).
\end{equation}
For an ideal Otto cycle, in order to avoid any complications arising from the irreversibility inherent to quantum friction , we shall consider perfectly adiabatic strokes by assuming an infinitely slow transformation time $t \to \infty$ , This regime is usually referred to as adiabatic, since the unitary evolution is sufficiently slow for the adiabatic theorem to hold, preventing transitions between the instantaneous eigenstates of the Hamiltonian , which leads us back to the Fermi–Dirac distribution;  \[
\langle \hat{b}_k^\dagger \hat{b}_k \rangle_{t \to \infty}  = \tilde{f}_k .
\]

The thermodynamic quantities are calculated as follows:
\begin{align}
&Q_h = \sum_k \omega_k(h_f) \Delta f_k ,\notag\\&
Q_c = - \sum_k \omega_k(h_i) \Delta f_k ,\notag\\&
W = \sum_k \left[ \omega_k(h_f) - \omega_k(h_i) \right] \Delta f_{k},\notag\\&
\eta = \frac{W}{Q_h}.
\end{align}
where we set 
\begin{equation}
\Delta f_k  \equiv f(\beta_h, \omega_k(h_f)) - f(\beta_c, \omega_k(h_i)).
\end{equation} 
To calculate the efficiency \( \eta_R \) of the refrigerator, defined as 
\begin{equation}
\eta_R = \frac{Q_c}{|W|}.
\end{equation}
\begin{figure*}[t]
    \centering
    \begin{subfigure}[t]{0.0\textwidth} 
        \centering
        \includegraphics[width=8.2cm]{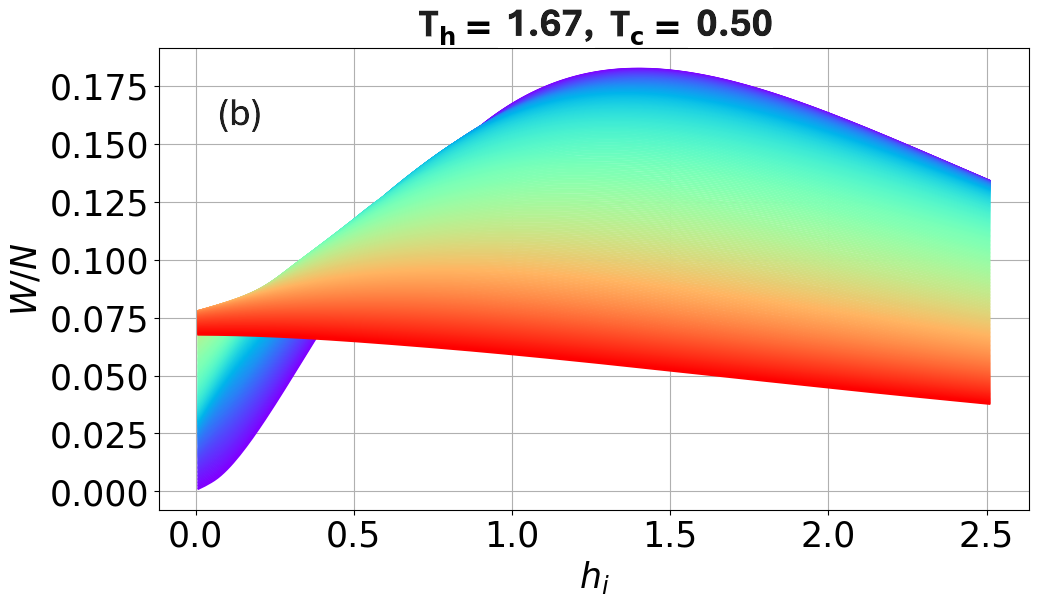} 
        
    \end{subfigure}
    \hfill
    \begin{subfigure}[t]{0.52\textwidth} 
        \centering
        \includegraphics[width=9.1cm]{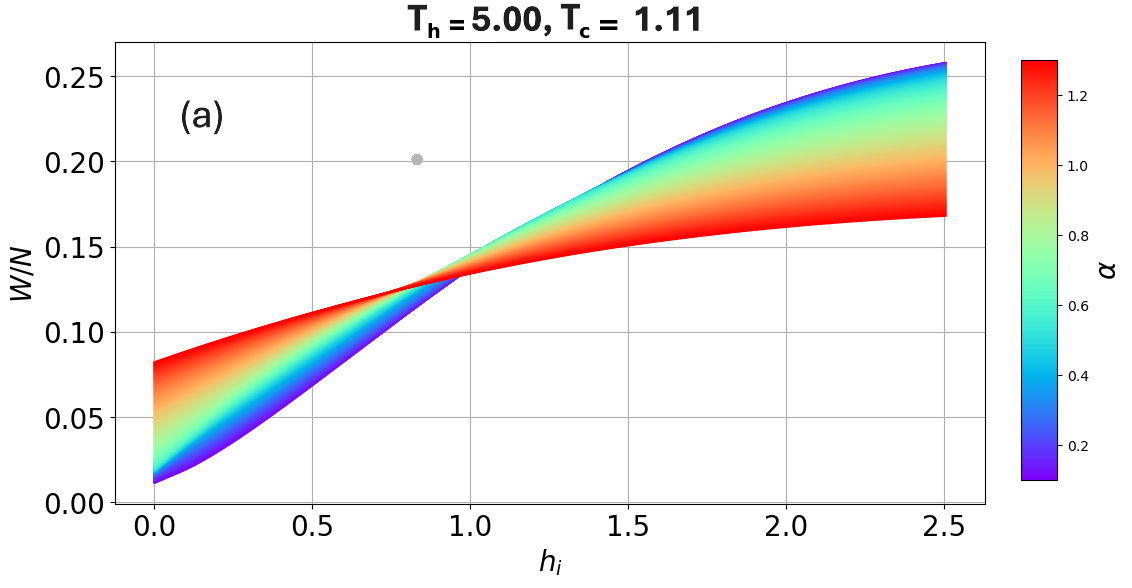} 
    \end{subfigure}

    \begin{subfigure}[t]{0.0\textwidth} 
        \centering
        \includegraphics[width=8.2cm]{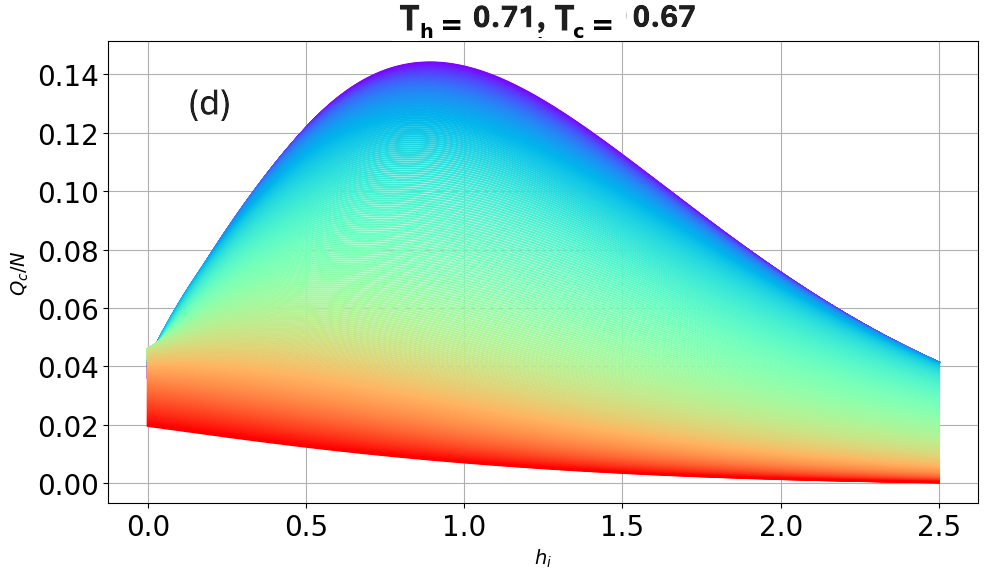}
       
    \end{subfigure}
    \hfill
    \begin{subfigure}[t]{0.52\textwidth} 
        \centering
        \includegraphics[width=9.2cm]{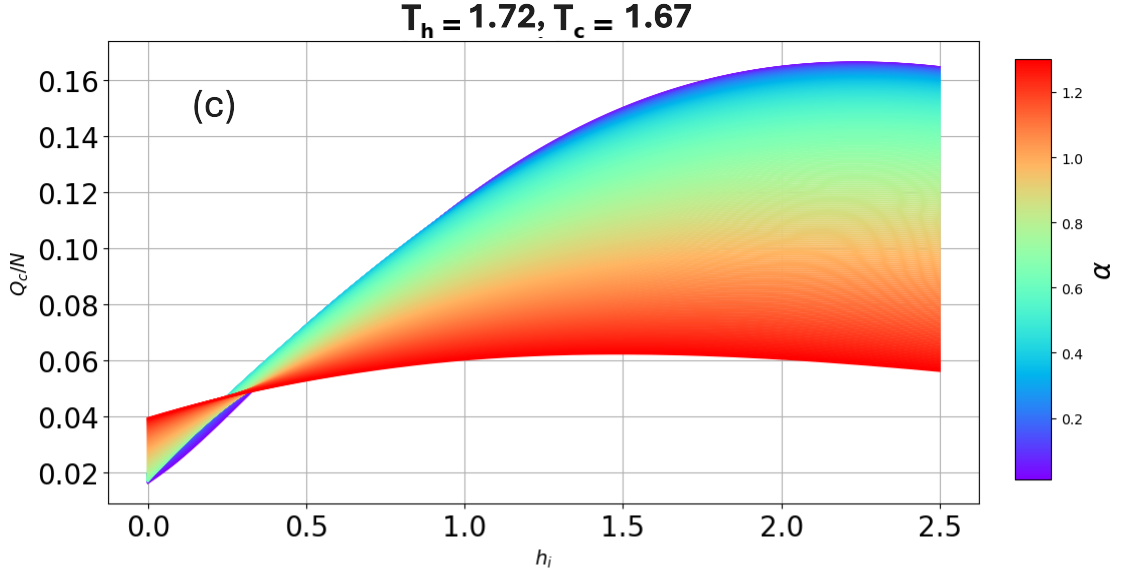}
    \end{subfigure}
 
    \caption{The influence of long-range interactions on the work and the quantum refrigerator's performance per spin as a function of \(h_i\). Here \(N\) is fixed at 100, while \(\alpha\) is varied from $0$ to $1.3$.}
\label{Fig77}
\end{figure*}

\begin{figure*}[t] 
    \centering
    \begin{subfigure}[t]{0.0\textwidth}
        \centering
        \includegraphics[width=8.0cm]{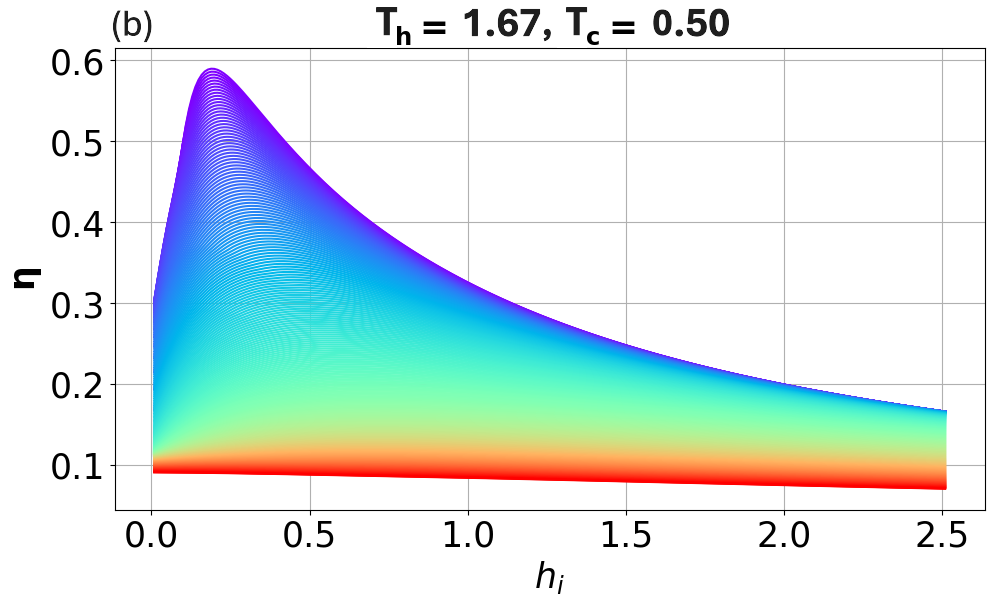}
    \end{subfigure}
    \hfill
    \begin{subfigure}[t]{0.54\textwidth}
        \centering
        \includegraphics[width=9.3cm]{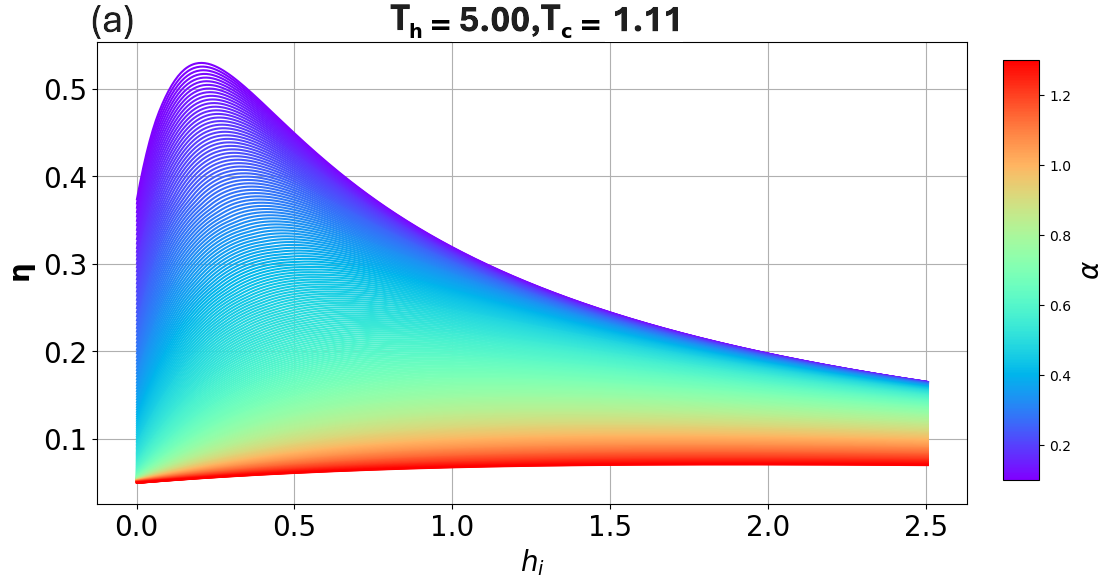} 

        \end{subfigure}

    \begin{subfigure}[t]{0.0\textwidth}
        \centering
        \includegraphics[width=8.0cm]{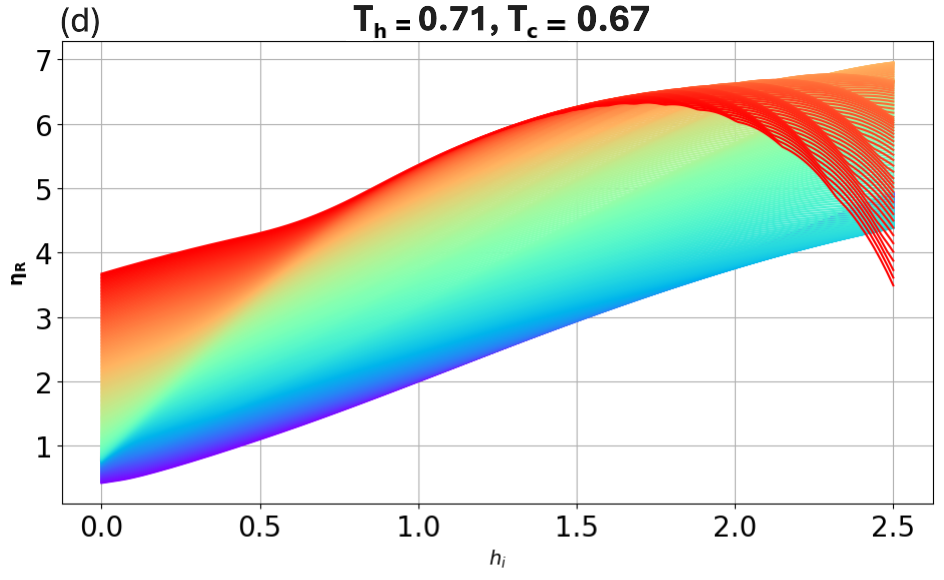} 
       
    \end{subfigure}
    \hfill
    \begin{subfigure}[t]{0.52\textwidth} 
        \centering
        \includegraphics[width=9.2cm]{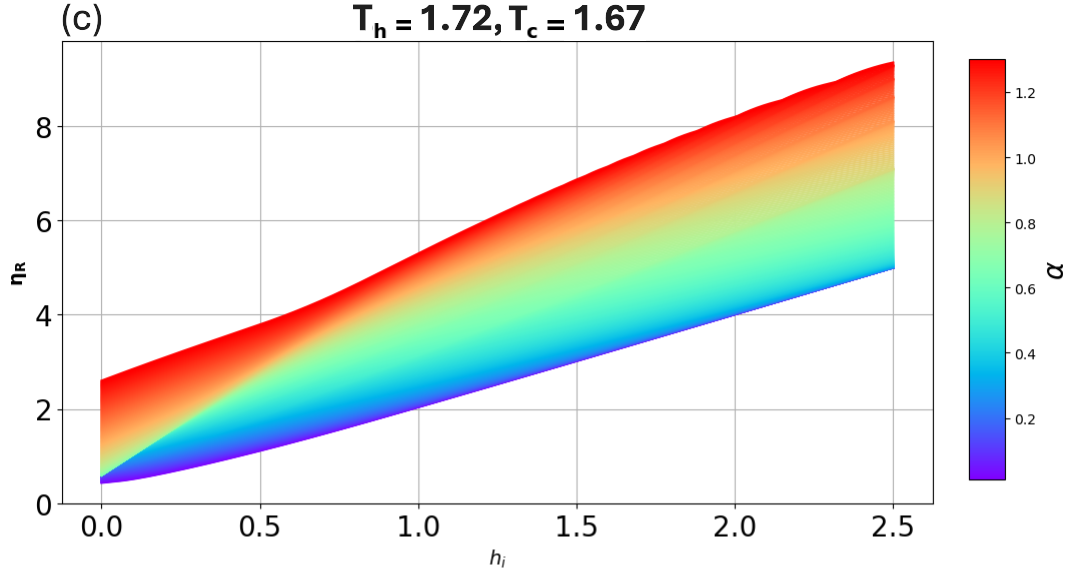}
    \end{subfigure}

    \caption{In the figure, \(N\) is fixed at 100, and \(\alpha\) is varied from 0 to 1.3 to observe the role of long-range interactions on the efficiency $\eta$ of the quantum heat engine and the coefficient of performance  $\eta_R$ of the refrigerator as a function of \(h_i\).}
 \label{fig5}
\end{figure*}
Note that, since after a single ideal cycle the system returns to the same initial state, we have $\Delta E = 0$ and thus $W = Q_h + Q_c$ follows from the first law of thermodynamics.

\subsection{Operations modes}

The purpose of this section is to show the switching between operational modes and their dependencies on the number of particles \(N\), the interaction strength \(\alpha\) , and the ratio \(\beta_h / \beta_c\).Let us now analyze the case of an infinitely slow cycle , the limit $t \to \infty$. All \textit{\(\alpha\)} values can be considered as long-range correlations, but for comparison purposes, we will divide them into three categories: long-range, intermediate, and short-range. The color code represents the following modes: accelerator (green), heater (purple), heat engine (red), and refrigerator (blue).As illustrated in Fig.\ref{fig2}, our engine can operate in one of the following four modes:

\begin{itemize}
    \item \textbf{Refrigerator (R)}: the engine absorbs energy and transfers heat from the cold reservoir to the hot one, i.e., $Q_c>0$, $Q_h<0$, and $W<0$;
    \item \textbf{Accelerator (A)}: the engine absorbs energy and transfers heat from the hot reservoir to the cold one, i.e., $Q_c<0$, $Q_h>0$, and $W<0$;
    \item \textbf{Heat engine (E)}: the engine produces work by absorbing heat from the hot reservoir, i.e., $Q_c<0$, $Q_h>0$, and $W>0$;
    \item \textbf{Heater (H)}: the engine absorbs energy and heats up both the hot and the cold reservoirs, i.e., $Q_c<0$, $Q_h< 0$,and $W<0$.
\end{itemize}

Figure~\ref{Fig3}-a summarises the machine's operational regimes for $N=100$, $T_h=0.57$ and $T_c=0.38$ as the interaction range is varied. Physically, decreasing $\alpha$ (making the interactions longer-ranged) strengthens collective correlations and coherent energy exchange across the chain: this favors refrigeration at low fields and promotes nonlocal response modes at intermediate fields, because the system can coordinate energy flows across many sites. By contrast, increasing $\alpha$ (shortening the interaction range) suppresses these collective channels, which reduces the region where refrigeration and transient modes appear and instead stabilises steady engine-like operation over a wider field window. In other words, longer-range couplings enhance sensitivity to global field conditions and produce pronounced finite-$N$ modes, whereas more local couplings favour robust, local energy extraction and smoother, more extensive engine performance.
The former case $\alpha=0.2$, lies well outside the perturbative domain of Fisher's expansion and corresponds to an interaction range so extended that non-additive, quasi--mean-field behaviour is expected; the latter, $\alpha=1.2$, remains within the classical long-range sector but with markedly weaker long-range effects. These conclusions are therefore fully compatible with the analytical expectations of \cite{v6} and with the qualitative differences observed in our numerical data.
Figure~\ref{Fig3}-b displays the machine's operational map for $N=100$ and $\alpha=0.25$, with $T_h=0.57$ while the cold temperature is varied so that the ratio $T_c/T_h$ spans $0.45$--$0.75$. Physically, changing $T_c/T_h$ tunes the thermodynamic bias available for work extraction: larger ratios (temperatures closer together) reduce the net thermal gradient and suppress steady work output, making thermally-driven refrigeration and transient response modes more prominent near the critical field. Conversely, decreasing $T_c/T_h$ (increasing the thermal contrast) strengthens the directional heat flow that drives net work production, stabilising engine-like operation across a broader field range and diminishing the parameter window where refrigerator or heater modes appear. Transient accelerator/heater bands that flank the critical region reflect a competition between critical susceptibility (which amplifies small driving imbalances) and the global thermal bias; as the thermal bias grows, this competition is resolved in favour of persistent engine behaviour.
Figure~\ref{Fig3}-c shows how the machines operating regimes reorganize as the system size $N$ is varied (here $\alpha=0.25$, $T_h=0.57$, $T_c=0.38$). Physically, the observed changes reflect a competition between two size-dependent effects: (i) finite-$N$ coherence and level discreteness, which favour directed, coherent energy extraction when the chain is short, and (ii) the progressive densification of the many-body spectrum and the growth of low-energy collective modes as $N$ increases, which amplify critical susceptibility and thermal redistribution near the critical field. For small chains, the sparsity of excitations and enhanced coherence make steady engine-like operation energetically favourable at low fields. As $N$ grows the spectrum becomes denser and low-energy fluctuations proliferate; these modes couple more efficiently to the reservoirs and to field-induced critical responses, expanding refrigeration and transient accelerator/heater bands close to criticality and reducing the parameter window for steady engine operation. At the largest sizes the increased phase-space for scattering and thermal equilibration further reinforces refrigeration-dominated behaviour over a broad field interval. In short, increasing $N$ shifts the device from a regime dominated by finite-size coherent extraction to one governed by many-body critical fluctuations and thermal redistribution, which explains the progressive replacement of engine by refrigerator behaviour in the data.\noindent The device's operational mode is set by three parameters: the interaction exponent $\alpha$, the thermal contrast $T_c/T_h$ (or $\Delta T$), and the system size $N$. Smaller $\alpha$ (long-range couplings) enhances global coherence and favours refrigeration with pronounced finite-$N$ transients near criticality, whereas larger $\alpha$ suppresses collective channels and stabilises engine-like operation. A small thermal gap (high $T_c/T_h$) weakens net work extraction and biases the system toward refrigeration or transient responses, while a large gap strengthens sustained engine performance. Small $N$ permits coherent finite-size windows for directed extraction; large $N$ increases low-energy fluctuations and thermalisation phase-space, broadening refrigeration and transient regimes. Hence $\alpha$, $T_c/T_h$, and $N$ constitute direct, tunable levers for steering and optimising quantum thermal machines.

\begin{figure*}[t] 
        \includegraphics[width=4.35cm]{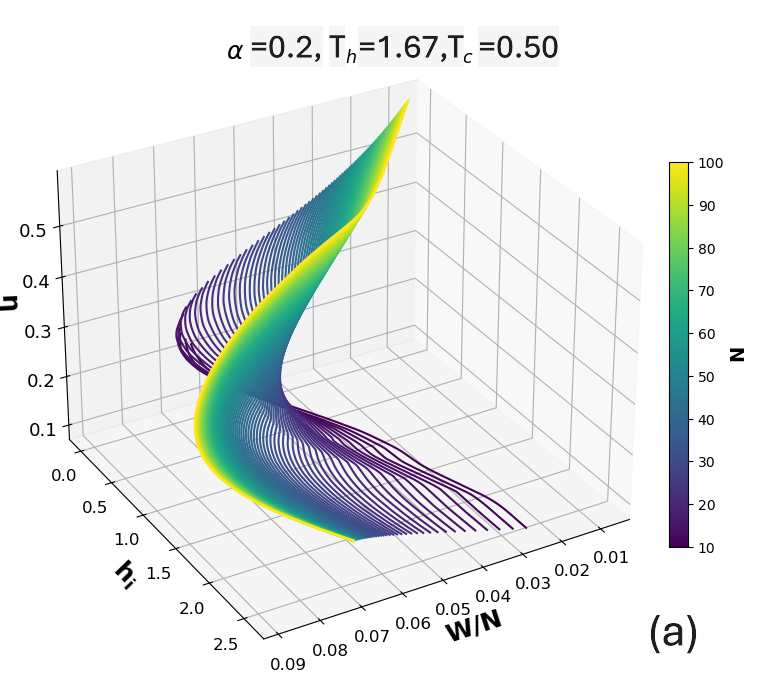}
        \includegraphics[width=4.35cm]{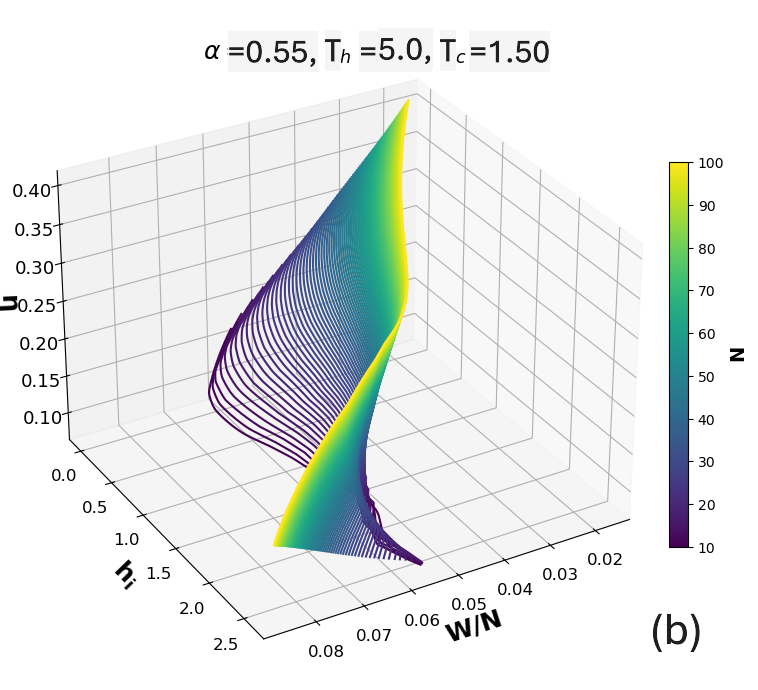}
        \includegraphics[width=4.35cm]{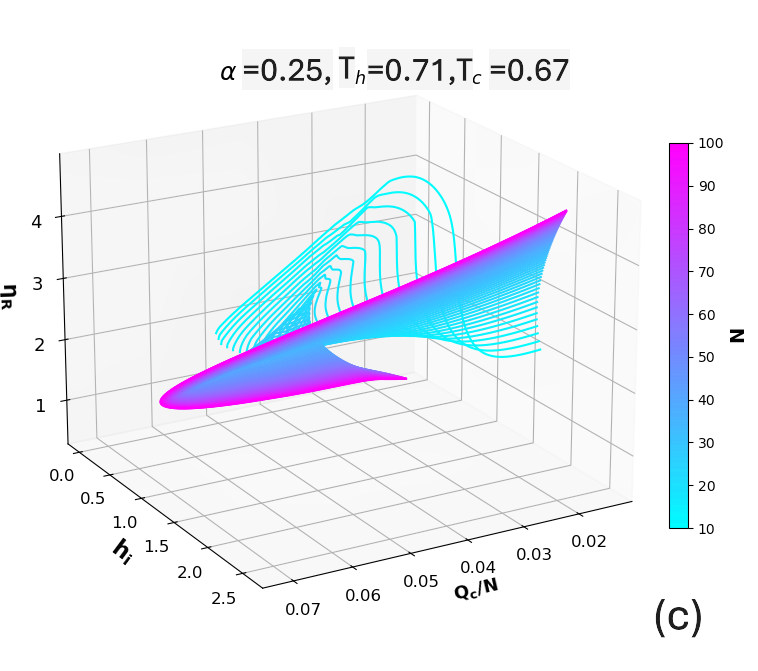}
        \includegraphics[width=4.35cm]{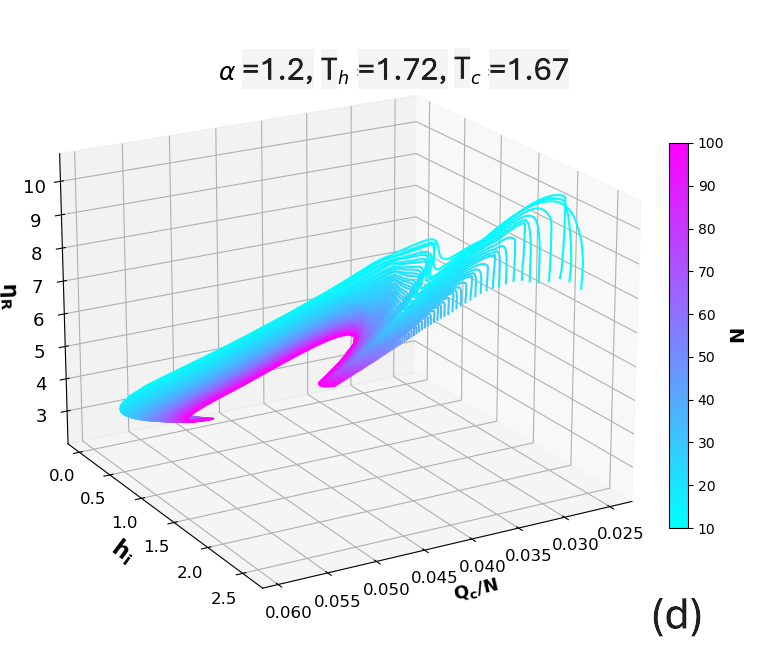}
    \caption{Efficient $\eta$ (COP $\eta_R$ ) as a function of $h_i$ and W/N ($Q_c$/N) this highlights the behavior of the efficiency for different initial field values \( h_i \) and of the work per spin \( W/N \), as well as the possible optimization paths of the efficiency for various interaction ranges characterized by different values of \( \alpha \).
.}
    \label{Fig5}
\end{figure*}

\subsection{Long Range Impact}

In the present study, we explore how the distinctive attributes of long-range interacting quantum systems can enhance the efficiency of a many-body quantum Otto cycle, adopted here as a representative model of a quantum thermal device. Our analysis reveals several key benefits similar to \cite{Solfanelli2020},  derived from the incorporation of long-range interactions:

\begin{enumerate}
    \item The infinitely slow limit , the device  attains markedly improved optimal performance. This enhancement is most evident in the practically relevant heat-engine and refrigerator regimes when collective and local effects are optimally balanced.
 .
    \item The interaction-range exponent $\alpha$ affects the cycle efficiency differently depending on the operational mode.

\end{enumerate}

Relying on the results reported by \cite{vv},
it was  shown that the spatial decay of connected correlations depends sensitively on $\alpha$ and on proximity to the critical  phase transition: large $\alpha$ yields short-range exponential screening, small $\alpha$ produces algebraic long-range tails, and an intermediate hybrid crossover combines both behaviours. 
On the paramagnetic side, quasi-exponential screening amplifies nearest-neighbour influence, whereas sufficiently small $\alpha$ restores system-wide coherence and collective channels. Crucially, the presence of a spectral gap does not preclude algebraic tails, so that spectral structure intrinsically limits the efficacy of nonlocal propagation. In Figures~\ref{Fig77} and~\ref{fig5}, thermodynamic performance is plotted as a function of the initial magnetic field $h_i$ for various values of $\alpha$.
When the temperature gap is large ($T_h = 5.0$, $T_c = 1.11$) , an increase in $\alpha$, corresponding to a reduction in the range of correlations, leads to a higher work output per spin before the critical region. At the critical phase, an inversion occurs where long-range interactions with smaller $\alpha$ outperform those with larger $\alpha$, suggesting that post-critical dynamics favor long-range interactions. This is explained by the influence of neighboring particles, which is more effective in the paramagnetic phase, where increased spin misalignment enhances work extraction. When the thermal gap is reduced (e.g., $T_c = 0.5$, $T_h = 1.67$), long-range interactions yield a higher $W/N$, as an increase in $\alpha$ induces localized interactions that propagate more weakly than long-range interactions when the thermal gap is small. The efficiency $\eta$ exhibits a similar trend across configurations, reaching a maximum for long-range correlations at low $h_i$, with a slight shift in the peak depending on the thermal gap. Finally, in the refrigerator configuration with a narrow thermal gap ($T_h = 0.71$, $T_c = 0.67$ and $T_h = 1.72$, $T_c = 1.67$), a similar behavior is observed for the cooling power $Q_c/N$, with long-range interactions yielding superior performance, as the extended range of interactions facilitates more effective heat diffusion compared to larger $\alpha$ values, which prioritize localized transmission. An increase in $\alpha$ for the efficiency $\eta_R$, which reduces dissipation to nearest neighbors, significantly enhances the yield until it reaches a saturation threshold. Guided by the conclusions of \cite{v1,v5}, the observed saturation of the critical peaks is a direct consequence of finite-temperature rounding: the thermal length scale \(\xi_T\) bounds the spatial correlations, so that finite-size scaling applies only for \(N\lesssim\xi_T\), whereas for \(N\gg\xi_T\) would-be critical divergences are suppressed and instead manifest as extended plateaus in the relevant observables. Long-range couplings (small \(\alpha\)) thus enhance performance for short chains but produce rapid saturation once \(N\) exceeds \(\xi_T\); by contrast, larger \(\alpha\) yields a more gradual increase in performance that ultimately levels off. Namely the coexistence in the QLRO phase of violations of the area law for entanglement entropy and a gapped entanglement spectrum are consistent with a picture in which the spread of quantum correlations is limited. This limited spread explains the saturation of the performance peaks we observe and suggests the existence of an energetic barrier that naturally limits the spread of quantum correlations.

\begin{figure*}[t]
        \includegraphics[width=4.3cm]{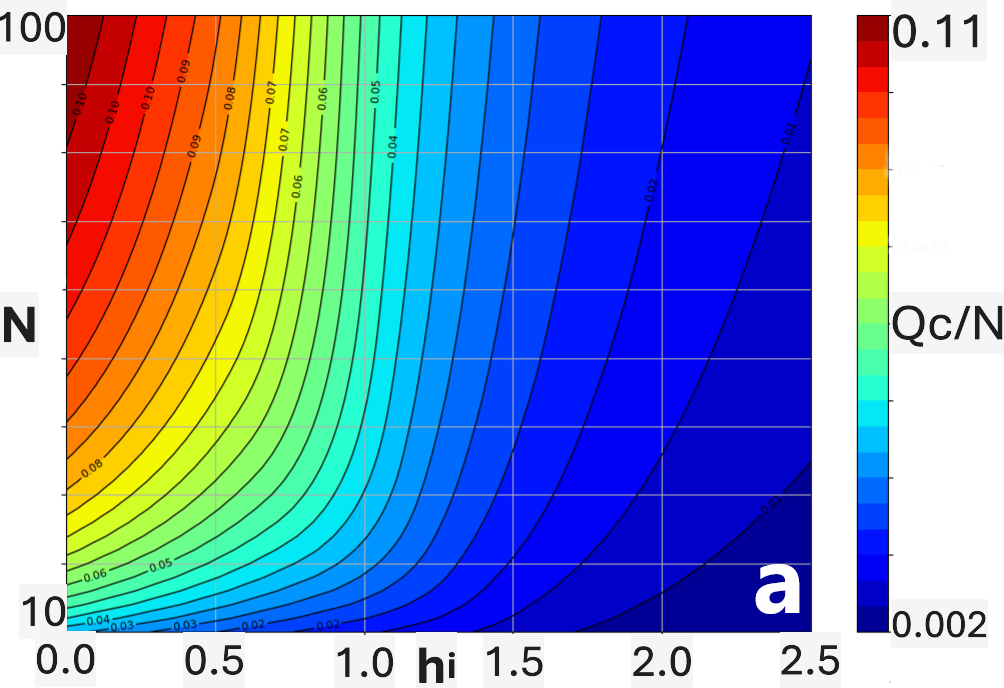} 
        \includegraphics[width=4.3cm]{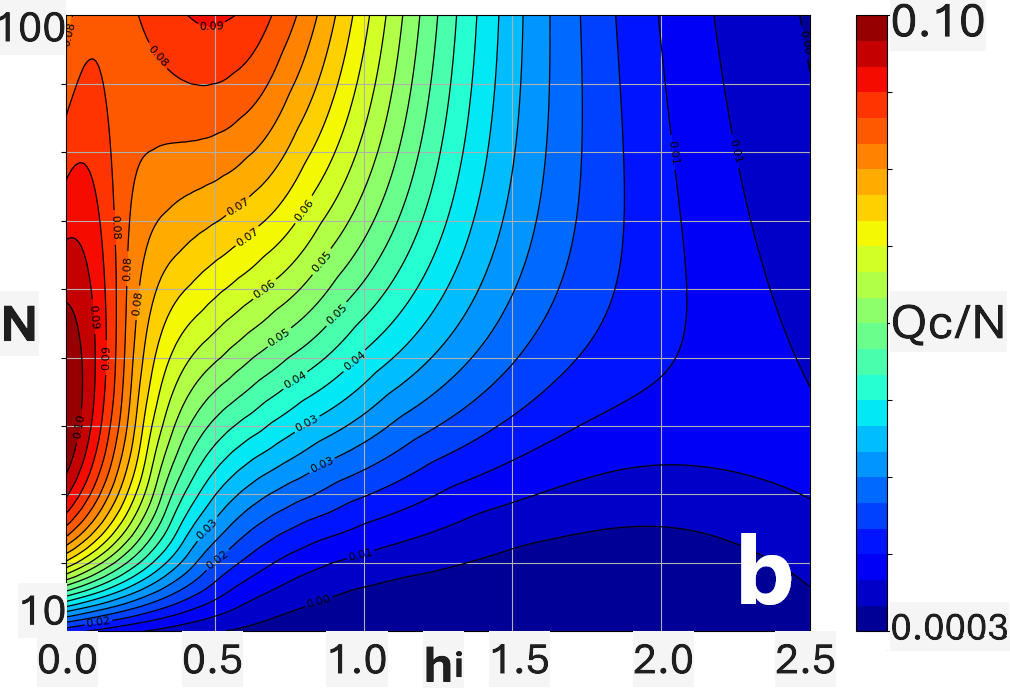} 
        \includegraphics[width=4.2cm]{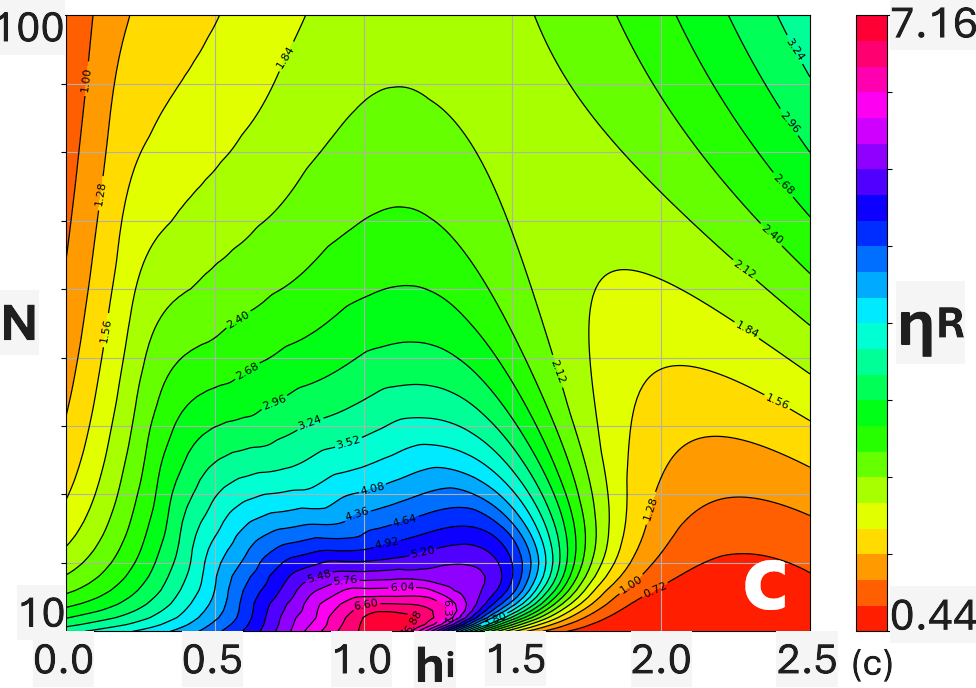}
        \includegraphics[width=4.2cm]{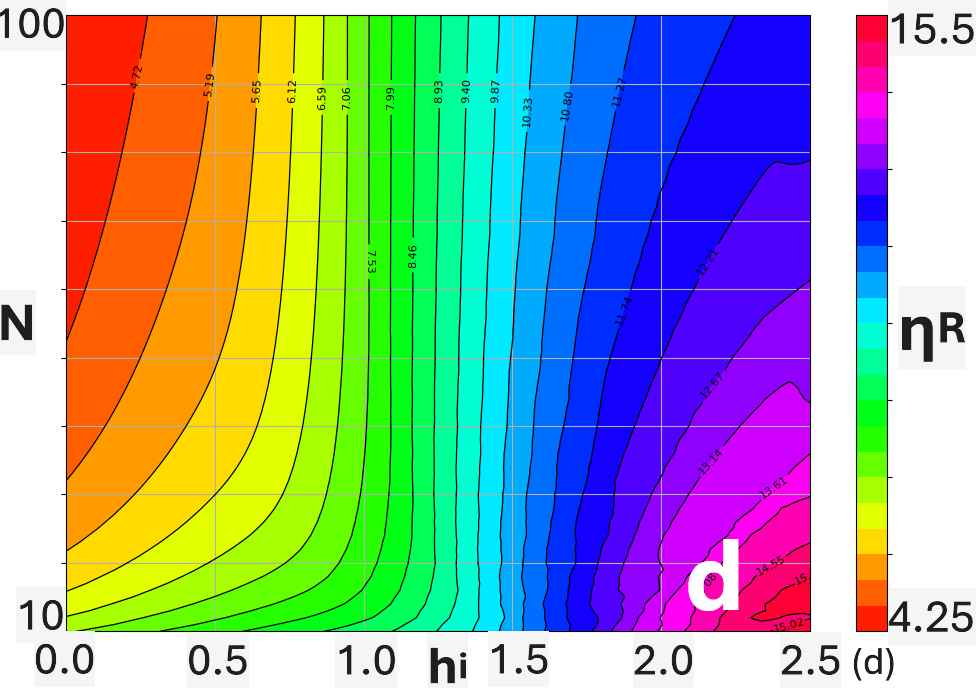}
    \caption{Contour plot of $Q_c$ denotes the the heat exchanged during the thermalisation
with the cold  bath done by the system and $\eta_R$ as a function of $ h_i$  for different values of \( \alpha \), with (a) \( \alpha = 0.25 \), (b) \( \alpha = 1.2 \), (c) \( \alpha = 0.25 \), and (d) \( \alpha = 1.2 \).}
    \label{Fig6}
\end{figure*}

\begin{figure*}[t] 
        \includegraphics[width=8cm]{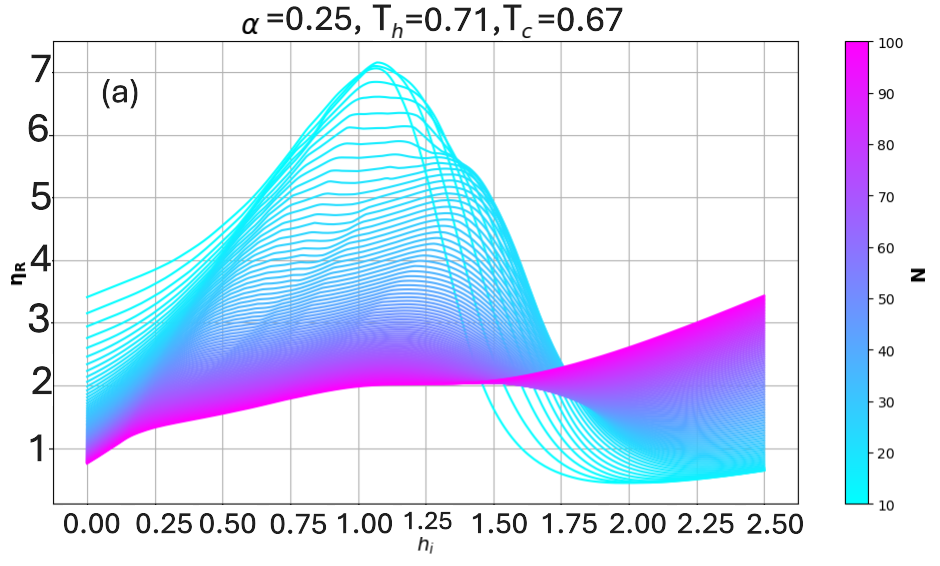}
    \includegraphics[width=8cm]{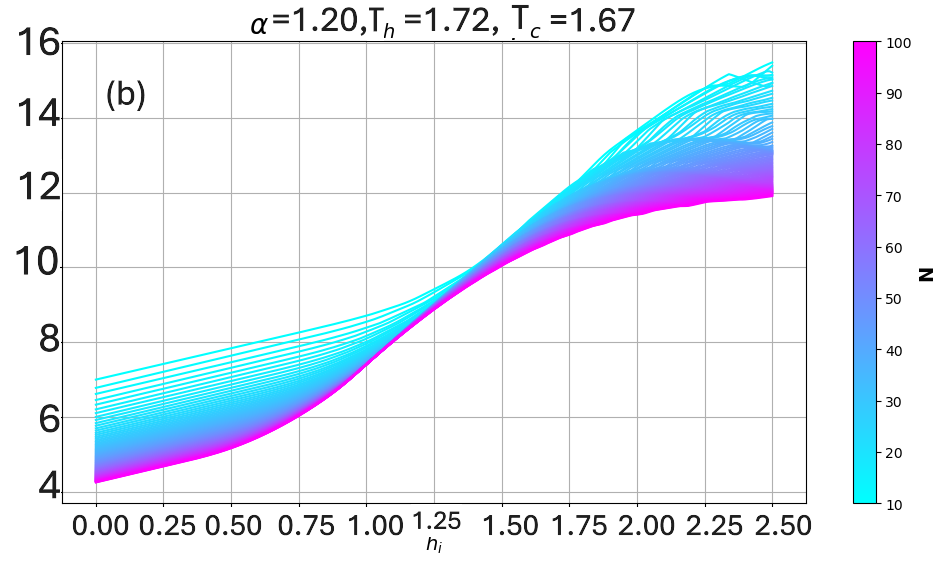}
    \caption{COP of the refrigerator versus $h_{i}$, for different system sizes $N$ across the critical point. The critical field is located at \( h_i = 1 \) for \( \alpha = 0.25 \), and approximately at \( h_i \approx 0.75 \) for \( \alpha = 1.2 \).
 }
    \label{Fig7}
\end{figure*}

To better understand optimization pathways, the behavior of thermodynamic efficiency $\eta$ and work per spin $W/N$ is analyzed as a function of the transverse field $h_i$ for different $N$ in Fig.~\ref{Fig5}.a-b. For small $N$, high values of $\eta$ are observed at low $h_i$, despite a higher $W/N$ compared to larger systems. Near the critical point, $h_i = 1$, short chains experience a sharp drop in efficiency, whereas longer chains exhibit a more gradual transition with a marked increase in $W/N$, displaying a more linear dependence on $N$. This contrasts with the refrigerator mode in Fig.~\ref{Fig5}.c-d, where short chains yield a higher $\eta_R$, while longer chains extract more heat. Post-phase transition, longer chains predominate due to dominant short-range correlations, which reduce sensitivity to phase transitions and thermal fluctuations. In summary, this work has investigated the performance of quantum heat engines and refrigerators by analyzing the role of interaction range—encoded in the parameter \( \alpha \)—across different thermodynamic regimes.

\section{Quantum heat engine and Refrigerator}
In this section, we will start by building a quantum refrigerator using our system and analyzing the behavior of physical quantities near the critical field. We will also investigate the effect of neighboring particles on the stability of the cycle through the scaling factor. Next, we will conduct a similar study on a quantum heat engine.
\begin{figure*}[t] 
        \includegraphics[width=8cm]{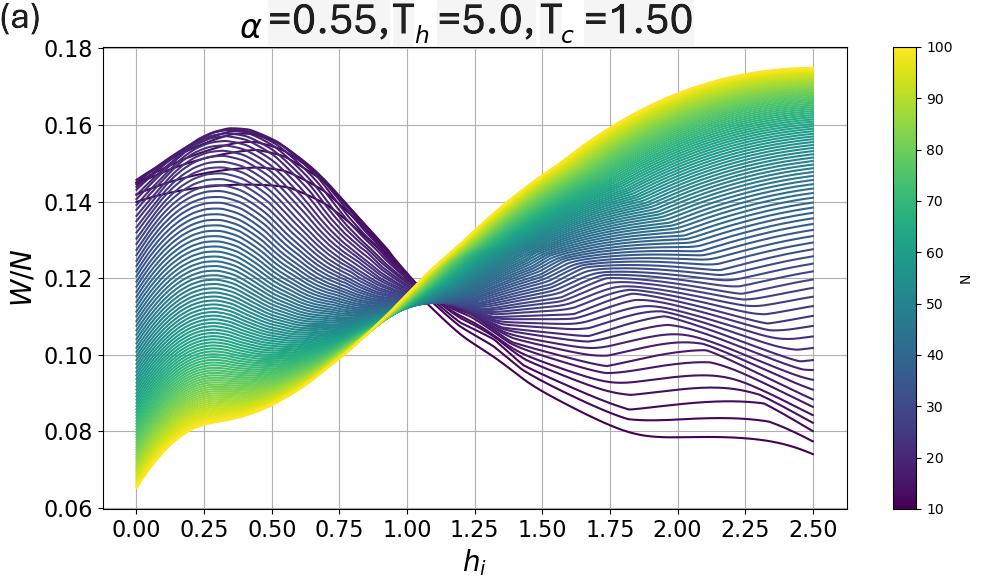} 
        \includegraphics[width=8cm]{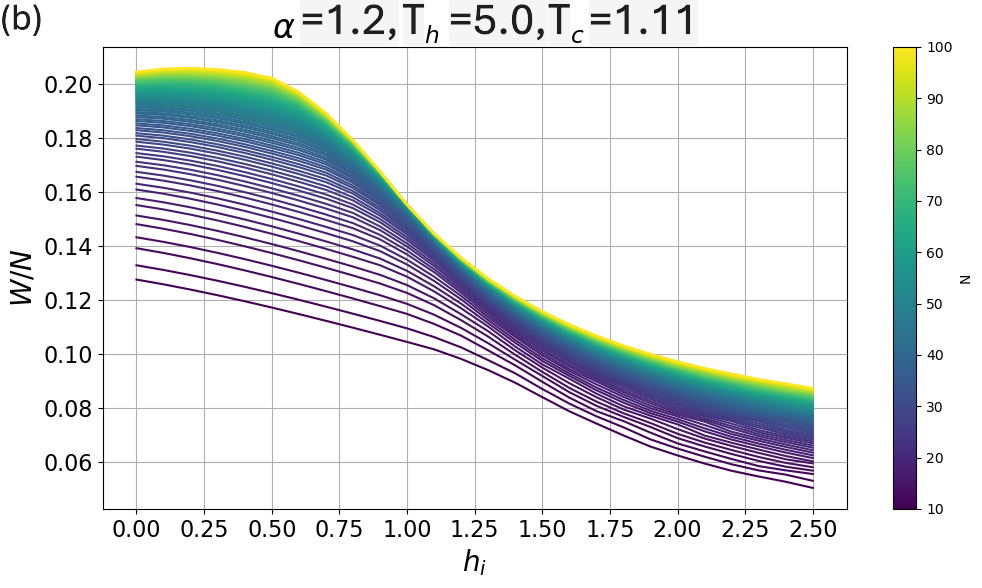}

        \includegraphics[width=8.0cm]{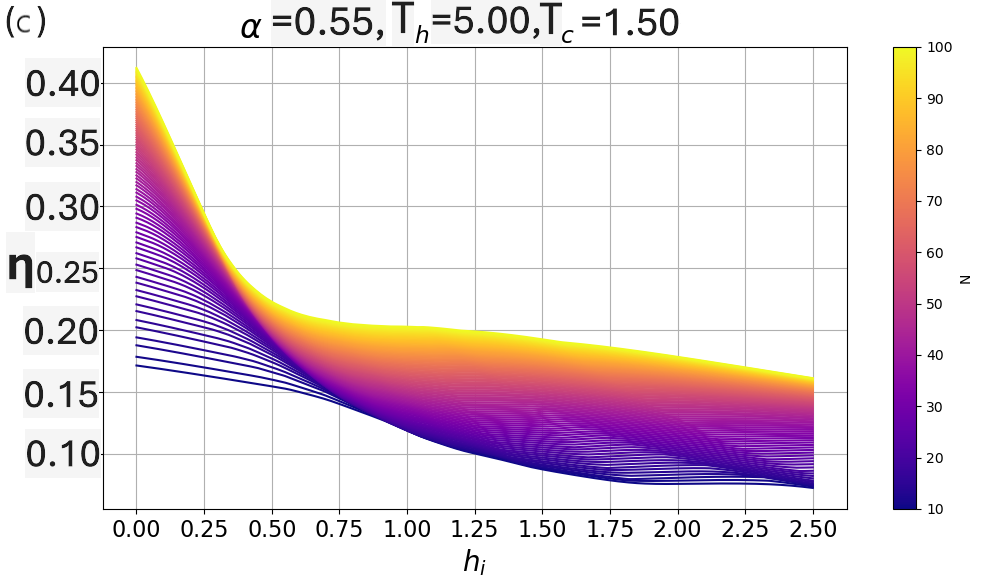}         \includegraphics[width=8.0cm]{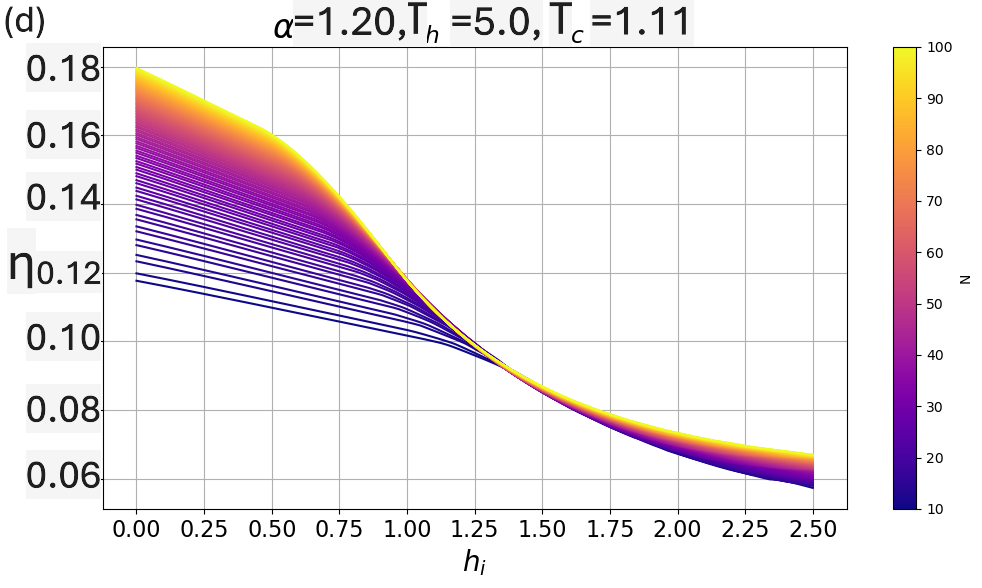} 
    \caption{Work per spin $W/N$ and efficiency $\eta$ as functions of the initial transverse field $h_i$ for different system size. For quenches performed in the paramagnetic phase, i.e., the critical field is located at \( h_i = 1 \) for \( \alpha = 0.55 \), and approximately at \( h_i \approx 0.75 \) for \( \alpha = 1.2 \).
for \( h_i > h_{\text{crit}} \), a distinct structure becomes apparent. In particular, the first peak, which becomes more pronounced as \( T_c \) increases, exhibits a superextensive scaling of the work with the system size \( N \), suggesting a potential influence of quantum critical behavior, same analysis as in the top panels, now applied to the efficiency \( \eta \) of the heat engine. The qualitative features remain consistent, including the dependence on \( N \) near the critical peak.
}
    \label{Fig10}
\end{figure*}

\subsection{ Quantum Refrigerator}

A thermodynamic advantage is  expected in the vicinity of the phase transition as highlighted in \cite{3d}, in finite-temperature systems near criticality, enhanced heat extraction is anticipated around the critical point. This enhancement arises from the divergence of the specific heat \cite{3p}, which allows substantial heat exchange even under small temperature gradients. A comparable reasoning applies to our magnetic system: as the quantum critical point is approached, the magnetic susceptibility—defined as the derivative of magnetization with respect to the external field—diverges. Consequently, the magnetization becomes highly responsive to even minor variations in the field. Since the work output due to a magnetic field change is proportional to the magnetization \cite{3e,3f}, this leads us to expect an increase in work extraction in the vicinity of the quantum critical point , similarly, improved heat extraction is expected in the refrigerator mode.
Our findings, as depicted in Fig.~\ref{Fig6}.a-b, demonstrate that for both values of $\alpha$, an increase in the interaction range enhances the strength of spin correlations and mitigates perturbations induced by neighboring particles, particularly in the ferromagnetic phase ($h_i < h_c$). This fosters coherent spin alignments, thereby improving efficiency. We observe superior heat extraction per spin, $Q_c/N$, for long-range interactions, which facilitate more effective energy transfer among nearest neighbors compared to short-range interactions that propagate more locally, thus minimizing $Q_c/N$. Long-range interactions lead to a smoother and more gradual decline in $Q_c/N$ beyond the critical transition, especially for longer chains, which exhibit greater stability and resilience. Conversely, short-range interactions result in a sharper drop around the critical point, attributable to the localized nature of energy diffusion compared to the more extended correlations observed for smaller $\alpha$. Drawing on the evidence provided by \cite{v2}, the authors showed that taxonomy separates three regimes by $\alpha$. At $\alpha=0.2$ (very long range, effectively near all-to-all) the system belongs to the non-additive, strong-long-range sector: without Kac rescaling the interaction energy grows super-extensively; collective, mean-field–like correlations and long-lived quasi-stationary states appear, relaxation is anomalously slow, and pronounced finite-$N$ peaks or apparent saturations dominate observables. By contrast, at $\alpha=1.2$ the model lies in the crossover from weak long-range to short-range behaviour, where energy is extensive, critical exponents and finite-size corrections become largely $\alpha$-independent, and observables approach conventional short-range scaling smoothly.\par

The analysis of the coefficient of performance (COP) corroborates the previously elucidated behavior in Fig.~\ref{Fig7}.a-b, where an increase in $\alpha$ leads to an enhanced COP, particularly for short chains, which preferentially favor the refrigeration mode. In contrast, long-range interactions, under the influence of neighboring particles, amplify perturbations and spin disorder, a phenomenon that is accentuated in the paramagnetic phase for longer chains, thereby fueling their performance.
The contour plots of $Q_c/N$ (Fig.~\ref{Fig6}.a-b) and COP (Fig.~\ref{Fig6}.c-d) highlight size-dependent dynamics. As $\alpha$ increases, contour lines become densely packed near the transition for small $N$, particularly for long chains, where the behavior is less smooth and more abrupt due to the nature of interactions, which are not as diffusive as long range interaction.

\begin{figure*}[t] 
        \includegraphics[width=6cm]{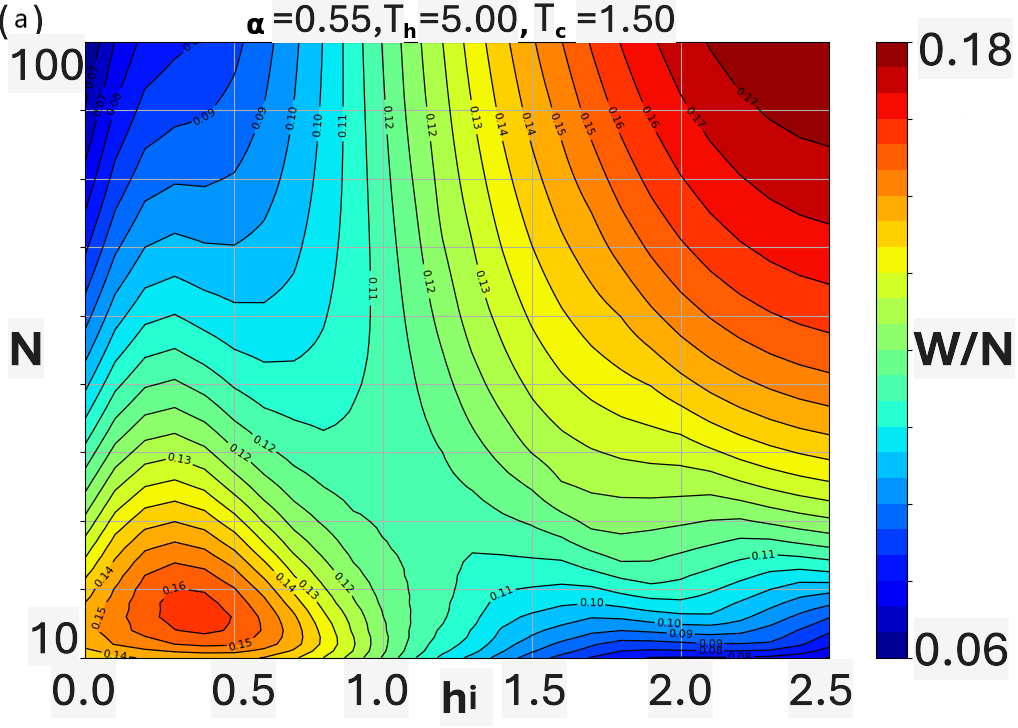} 
        \includegraphics[width=6cm]{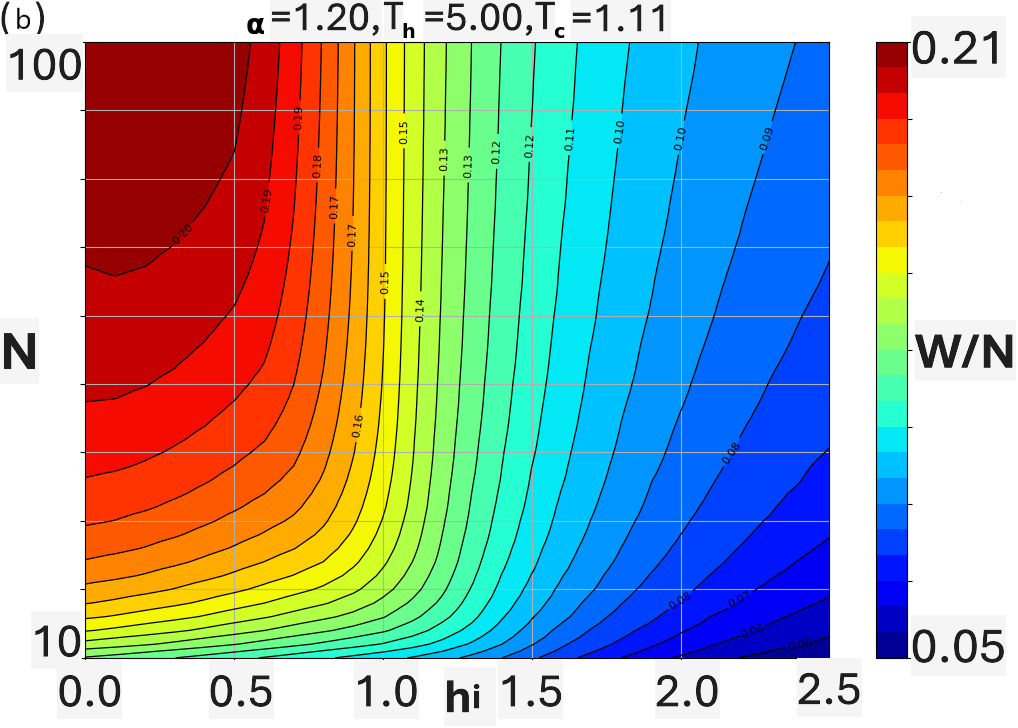} 

\includegraphics[width=6cm]{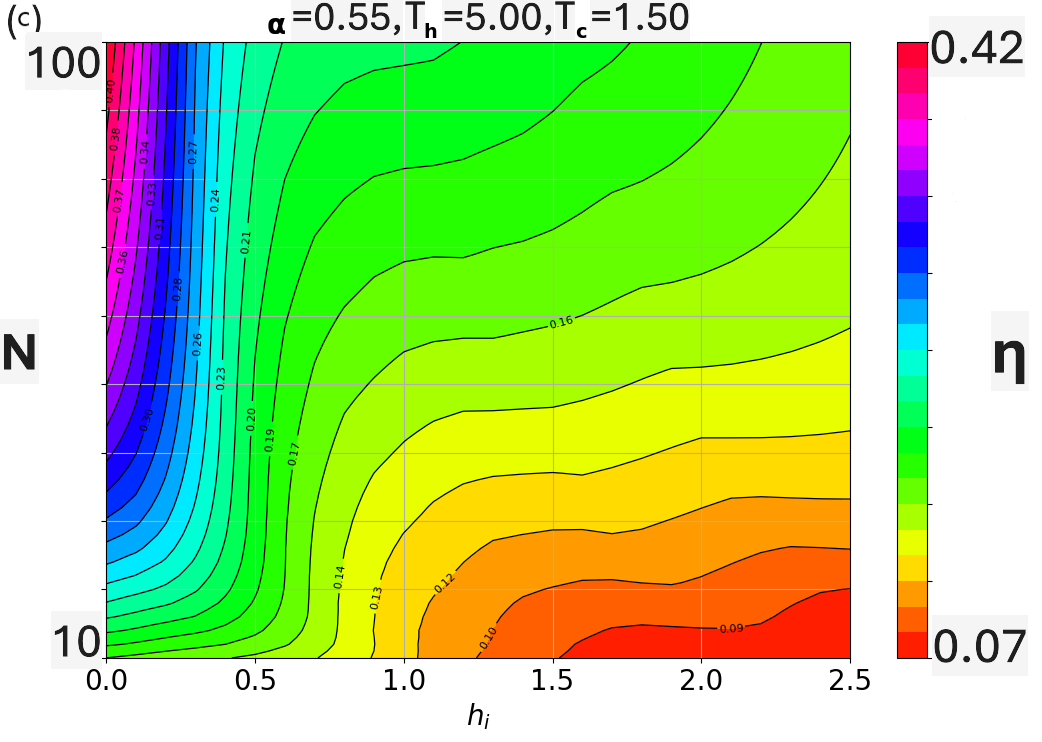} 
\includegraphics[width=6cm]{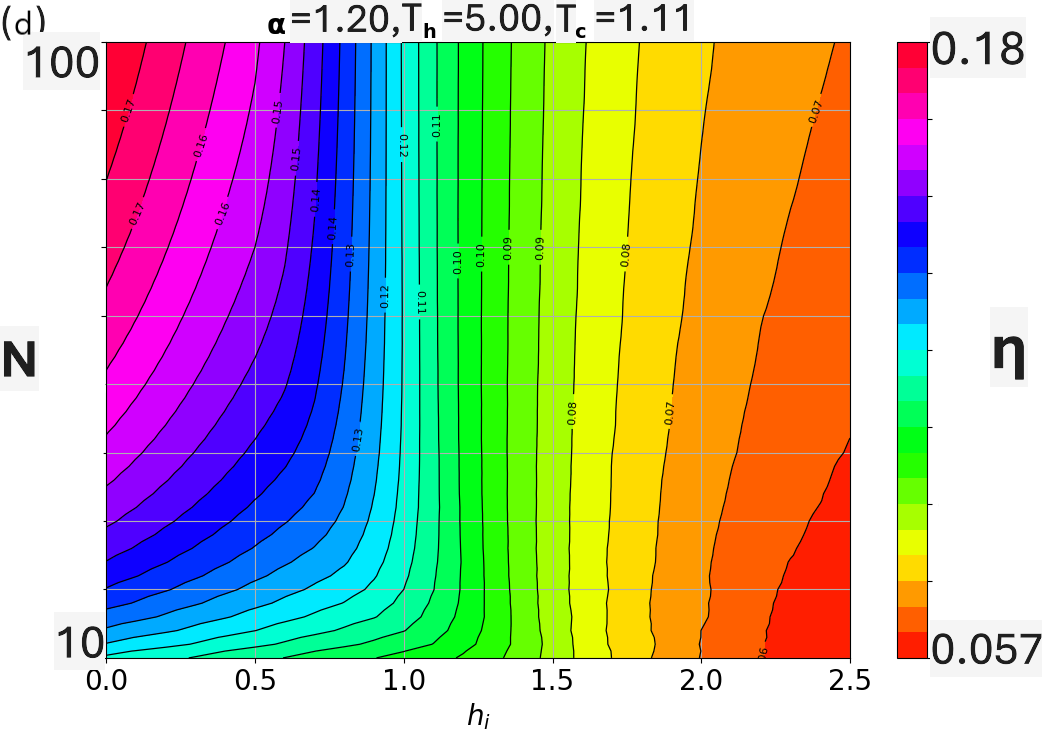} 
    \caption{     Contour plot of the work per spin $W/N$  and $\eta$ as a function of $ h_i$ for different sizes N , for different values of interaction range  \( \alpha \) . }
    \label{Fig11}
\end{figure*}

Overall, the observed deviations for small values of $N$ are indeed expected and can be attributed to well-documented finite-size effects, including discretized energy levels, boundary contributions, and the lack of a thermodynamic limit, cause fluctuations and non-extensive behaviors in quantities like heat and work per spin in small spin chains  \cite{3,4}. Quantum coherence and entanglement further influence the performance of quantum thermal machines at small $N$, leading to non-classical thermodynamic signatures. As $N$ increases, these effects diminish, and observables like work and heat per spin align with the extensivity principle of thermodynamics, consistent with known physics of finite-size systems \cite{5,6}. The system's behavior strongly depends on the chain size and the correlation regime, with each configuration offering specific advantages depending on the magnetic field. The performance per spin is better in the ferromagnetic phase, with efficiency varying depending on the type of correlation $\alpha$.

\subsection{Quantum Heat Engine}
We will see in this section the evolution of the work per spin ($W/N$)  and efficiency (\(\eta\)) as a function of the initial transverse field ($h_i$) for different values of $\alpha$, $T_h$, and $T_c$, analyzing the effects of the phase transition and the system's behavior for small and large chains ($N$). Each graph in   Fig.\ref{Fig11}  highlights how interactions and thermal conditions influence the system's performance.

Fig~\ref{Fig10}.a-b elucidates the behavior of the work per spin, \( W/N \), as a function of varying \(\alpha\), \( T_h \), and \( T_c \), highlighting the influence of chain length \( N \), transverse field \( h_i \), and differing interaction ranges. Increasing \(\alpha\) reduces the influence of neighboring particles while enhancing localized correlations, thereby promoting stronger spin alignment in the ferromagnetic phase (\( h_i < h_c \)). As illustrated in Fig~\ref{Fig10}-b, elevating \(\alpha\) generally leads to improved work extraction per spin, owing to diminished perturbations from proximate neighbors, resulting in a more linear behavior for large \( N \). This contrasts with the case of \(\alpha = 0.55\), where more pronounced perturbative effects from neighboring particles intensify with increasing \( N \), leading to a rapid misalignment of spins and consequently higher work extraction for smaller \( N \). In contrast, the paramagnetic phase (\( h_i > h_c \)), characterized by disordered spins, leverages these neighboring perturbations as a driving force, significantly enhancing work extraction, particularly for longer chains.              Regarding the efficiency, $ W/Q_h $, Figure~\ref{Fig10}c-d highlights distinct regimes where, in contrast to the work per spin, an increase in $\alpha$ diminishes the cycle’s efficiency. This is attributed to $ Q_h $, the heat absorbed from the hot reservoir, which is more effectively catalyzed and absorbed under long-range interactions. Conversely, at $\alpha = 1.2$, the more localized diffusion imposed by short-range interactions restricts this process, leading to reduced efficiency.Complementing this analysis, Fig.~\ref{Fig11}.a-b illustrates the contour lines of $W/N$, revealing dependencies on $\alpha$ and the thermal gap. For small $\alpha$, two peaks emerge for short and long chains in phases exhibiting distinct behaviors within the same interaction range, characterized by concentric circles centered around small $N$ in the ferromagnetic phase and large $N$ in the post-critical phase. For larger $\alpha$, a single peak at $\alpha = 1.2$ dominates in the post-critical phase for large chains, where the locality of short-range interactions introduces a linear dependence before the phase transition, followed by a gradual decline toward higher field strengths.

\begin{figure*}[t] 
        \includegraphics[width=7cm]{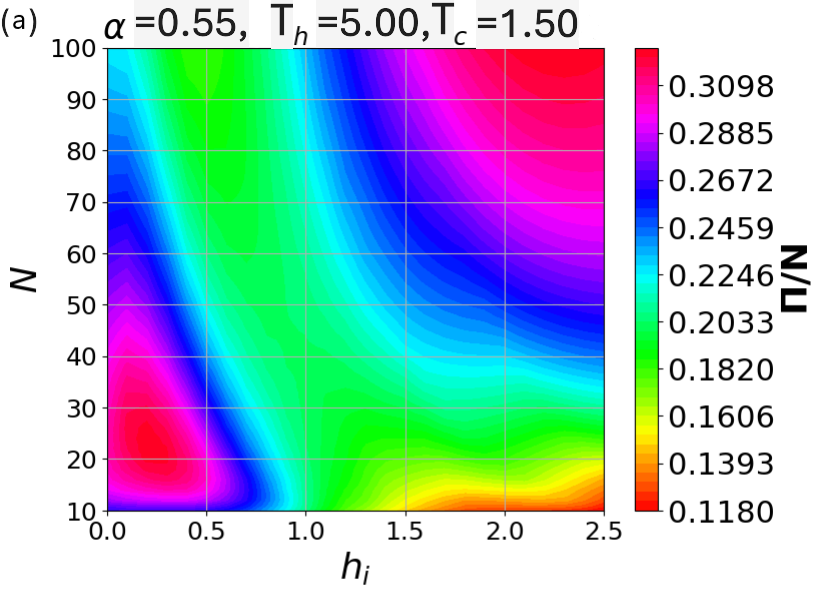} 
        \includegraphics[width=7cm]{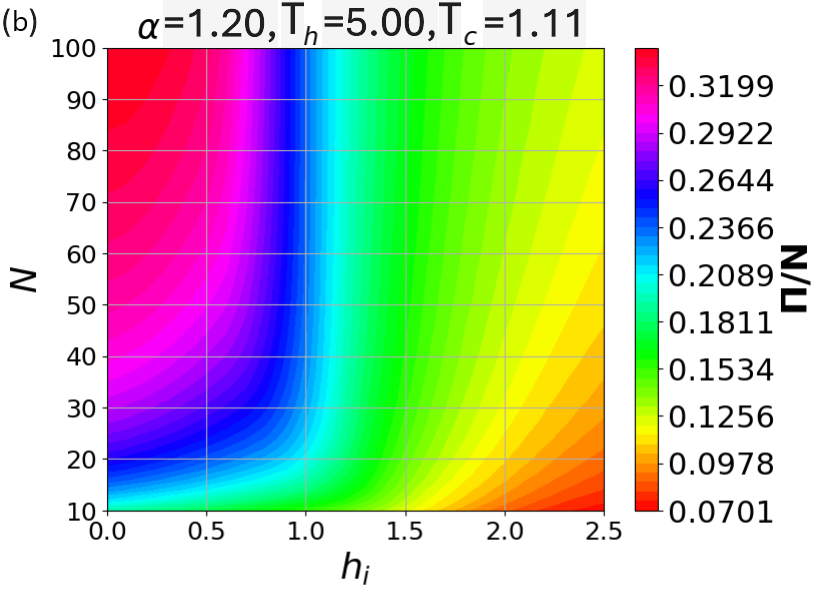} 
\includegraphics[width=7.2cm]{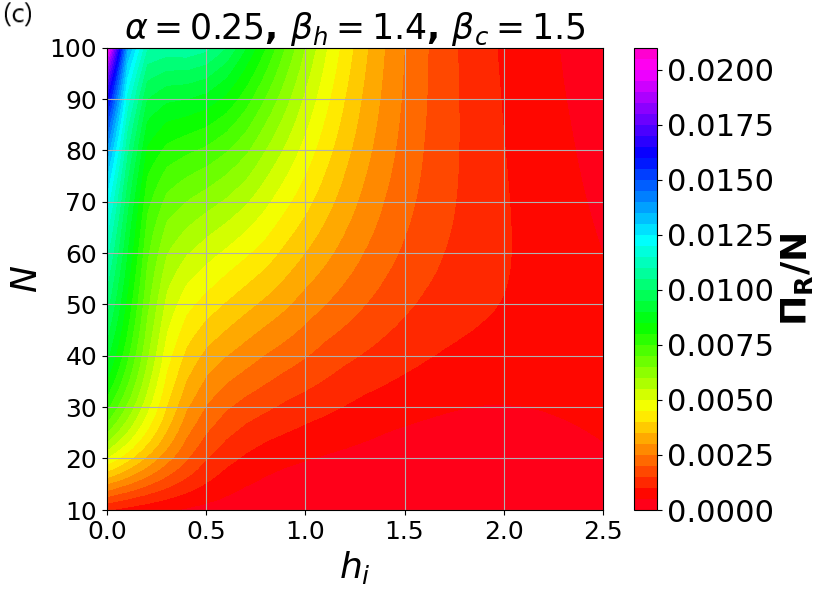} 
        \includegraphics[width=7.2cm]{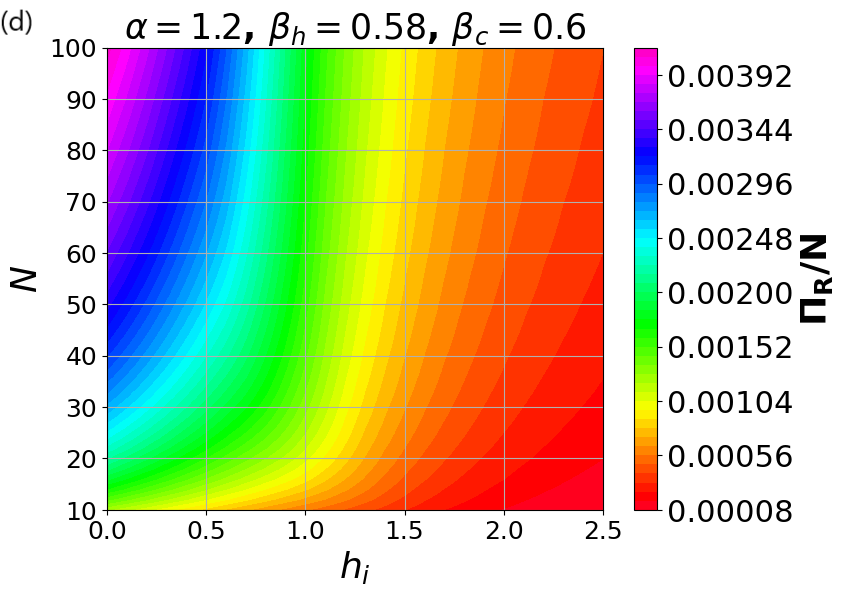} 
    \caption{The scaling factor per spin \( \Pi/N \) ( \( \Pi_R/N \)) as a function of \( h_i \), for various system sizes \( N \) and different interaction ranges characterized by \( \alpha \).}
    \label{Fig12}
\end{figure*}

Regarding efficiency, $W/Q_h$, Fig.~\ref{Fig11}.c-d highlights distinct regimes. For $\alpha = 0.55$, efficiency decreases gradually, with a rapid vertical drop for long chains before the phase transition, in contrast to $\alpha = 1.2$, which exhibits a more tempered decline and varies horizontally due to the linearity of short-range interactions. Thus, the relationship between efficiency and work per spin reflects the complex interplay of thermal fluctuations and correlations in the system, especially across different interaction ranges.

\subsection{Scaling factor of the quantum heat engine and refrigerator}

The scaling factor \( \Pi  \) measures how effectively an engine produces work despite having an efficiency lower than the Carnot limit \cite{Campisi2016}. Furthermore , an elevated $\Pi$ plays a pivotal role in criticality and phase transitions, where it indicates enhanced system stability against critical fluctuations and the capacity to sustain optimal performance during such transitions. Consequently, a high scaling factor reflects increased efficiency, resilience to perturbations, and optimal utilization of energy resources, even though the absolute efficiency remains below the theoretical maximum.To observe the impact and influence of correlations without considering the size of the chain, we will normalize by dividing by \( N \) , defined as :
\begin{equation}
\Pi /N = \frac{W}{(\eta_{\text{Carnot}} - \frac{W}{Q_h})} / N,
\end{equation}
where \(W\) is the work done by the system, \(\eta_{\text{Carnot}} = 1 - \frac{T_c}{T_h}\) is the Carnot efficiency, and \(N\) is the number of particles in the system in Fig.~\ref{Fig12} . These findings underscore the interplay of chain size, field strength, and interaction range in dictating system performance across phase transitions.  
Physically, $\Pi / N$ reflects the system's capacity to harness quantum correlations for work output, modulated by the interplay between $\alpha$ and the thermal gradient. Near criticality in finite-temperature systems, enhanced heat extraction is anticipated due to the divergence of specific heat, allowing significant heat exchange with minimal temperature gradients. Similarly, in this magnetic system, the magnetic susceptibility—diverges near the quantum critical point, making magnetization highly responsive to small field changes. Since work output from a magnetic field variation is proportional to magnetization , this heightened sensitivity boosts work extraction at criticality.An increase in $\alpha$ in Fig.~\ref{Fig12}-b enhances the strength of spin-spin correlations and reduces perturbations induced by neighboring particles, particularly in the ferromagnetic phase ($h_i < h_c$), promoting coherent spin alignments and elevating $\Pi / N$ by maximizing work extraction. Conversely, lower $\alpha$ values in Fig.~\ref{Fig12}-a  lead to long-range interactions with faster decay, increasing scattering and broadening the phase transition into a wider critical region . This renders short chains more stable before the phase transition. This diminishes the magnitude of $\Pi / N$ in the ferromagnetic phase due to weaker correlations but mitigates the impact of critical fluctuations, resulting in a smoother decline of $\Pi / N$ across $h_c$, particularly in smaller systems less sensitive to long-range effects . Let's move on to the refrigerator mode, we operate with $\Pi_R / N$, 
\begin{equation}
\Pi_R /N = \frac{Q_c}{(\eta_{\text{RCarnot}} - \frac{Q_c}{W})} / N,
\end{equation}
where \(Q_c\) denote the heat exchanged during the thermalisation
with the cold  bath done by the system, \(\eta_{\text{RCarnot}} =  \frac{1}{T_h /  T_c -1}\) is the Carnot refrigerator efficiency , which reflects the system's capacity to efficiently extract heat by leveraging spin correlations, modulated by $\alpha$ and the thermal gradient. Similarly to the engine mode, an increase in $\alpha$ influences the strength of correlations with neighboring spins (Fig.~\ref{Fig12}-c) .

\begin{figure*}[t] 
        \includegraphics[width=8cm]{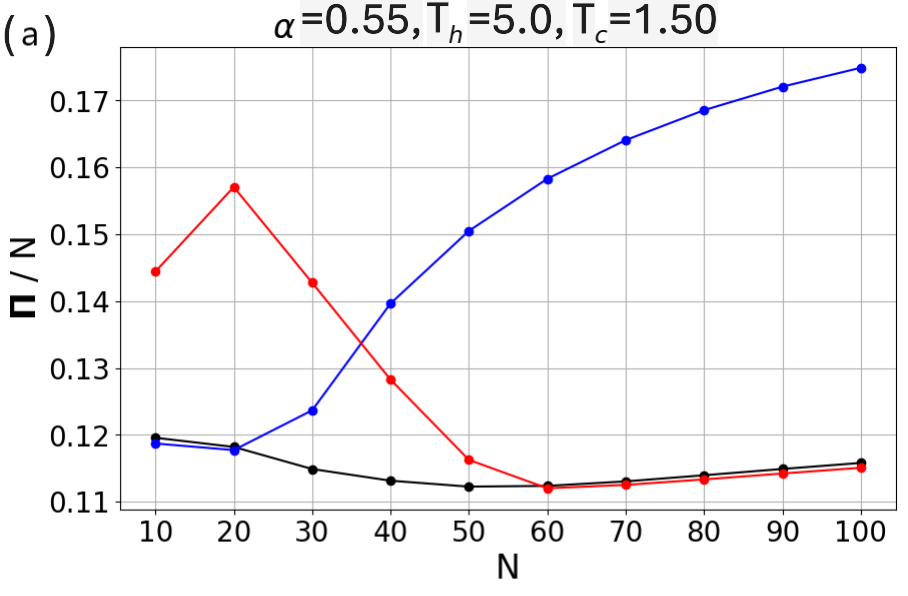} 
        \includegraphics[width=8cm]{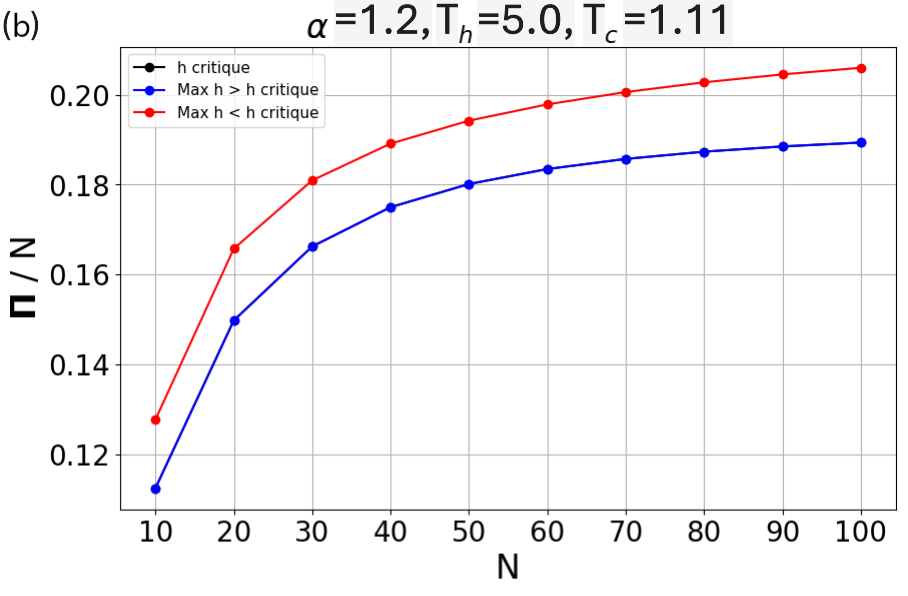} 
    \label{Fig9}
\end{figure*}
\begin{figure*}[ht] 
       
        \includegraphics[width=8.0cm]{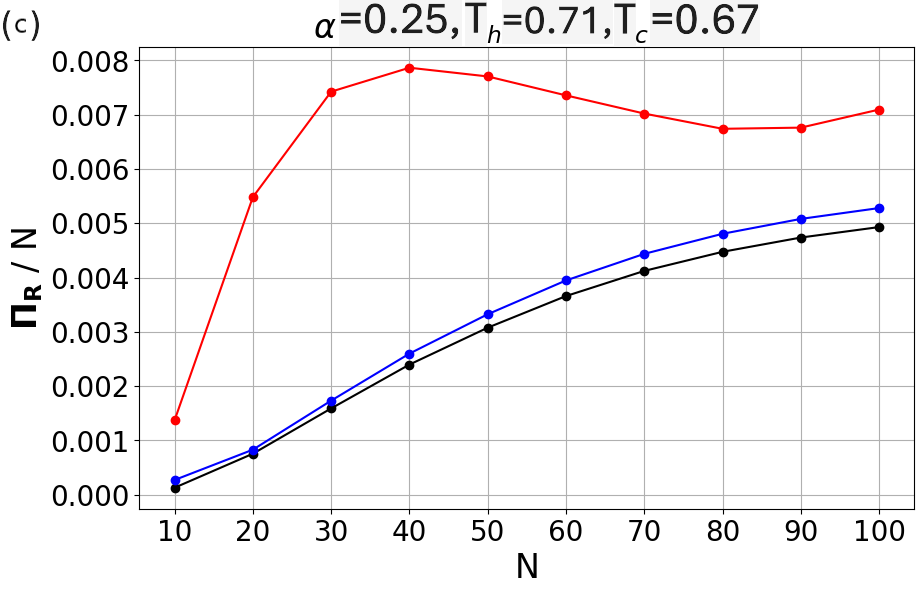}         \includegraphics[width=8.0cm]{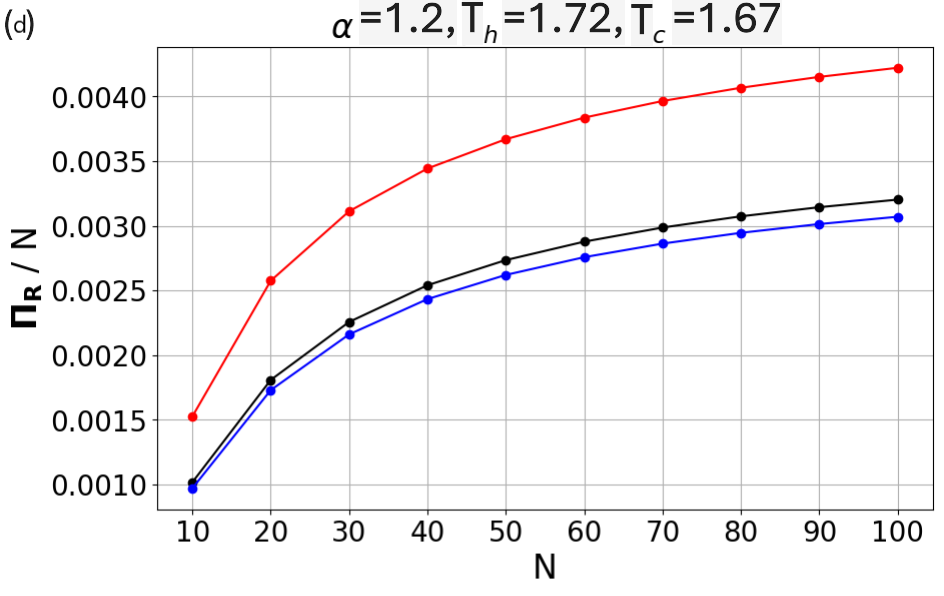} 
    \caption{The evolution of the maxima as a function of \( N \) (the number of particles) for the ferromagnetic peak (in red) and the paramagnetic peak (in blue) of \( \Pi/N \) and \( \Pi_R/N \), with a comparison to the critical field (in red).We will also analyze the peaks in the paramagnetic phase before the phase transition and the ferromagnetic phase after the transition, as well as their proportionality and dependence on N for the quantum heat engine and refrigerator mode by using  a regression to calculate the exponent \(a\) responsible for the growth rate of the scaling factor per spin, proportional to \(N^a\).}
    \label{Fig14}
\end{figure*}

However, unlike the engine mode, heat extraction from the reservoir is more effective when $\alpha$ is lower. This is explained by the divergence of the specific heat near the phase transition, which diminishes as $\alpha$ increases, resulting in a reduced potential for heat extraction and thus providing greater stability for long-range interactions.

In the presence of short-range correlations, post-transition performance declines, with a notable dependence on system size; longer chains demonstrate greater efficiency across all regimes. Additionally, for large \(\alpha\), short-range correlations prevail, bolstered by the thermal gap, which mitigates thermal fluctuations and enhances system efficiency. 
The behavior of \(\Pi/N\) as a function of \(h_i\) demonstrates the intricate relationship between work, system efficiency, and particle number \(N\). Large chains tend to dominate in regimes where long-range correlations are significant, especially beyond the critical point, whereas small chains are more effective in the short-range, pre-transition regime. 

\section{Investigating the role of thermal correlations and criticality}
In this section, we will examine the scaling factor \( \Pi/N \), studying its impact on the system's behavior for varying values of \( N \) (ranging from 10 to 100) and different configurations of \( \alpha \),  \( T_h \), and \( T_c \) .  focusing on the behavior near the critical field, including the peaks observed in the paramagnetic and ferromagnetic phases. Additionally, we will discuss how these features depend on \( N \) in both the quantum heat engine and refrigerator modes.

\subsection{Critical reaction}
We will examine the role of the scaling factor and the behavior at the critical field for $N$ ranging from 10 to 100. We will also analyze the peaks in the paramagnetic phase before the phase transition and the ferromagnetic phase after the transition, as well as their proportionality and dependence on N for the quantum heat engine and refrigerator mode. We intend to use a regression to calculate the exponent \(a\) responsible for the growth rate of the scaling factor per spin, proportional to \(N^a\). We will then examine the behavior of the performance and its trend for small, intermediate, and long chains and the role of thermal gaps and correlations $\alpha$ , to observe the impact of the thermal reservoirs on our system. The goal is to identify conditions under which increasing $N$ (number of spins or system size) leads to a significant improvement in yield compared to the energy cost. When $\alpha > 0$, it means that increasing $N$ results in a superlinear growth of cycle work and engine efficiency,  to achieve better performance on a large scale. 

The parameter $\alpha$ governs the decay of spin interactions, transitioning from a long-range regime ($\alpha \leq 1$) to a short-range regime ($\alpha > 1$). At a low $\alpha = 0.2$ ($T_h = 1.67$, $T_c = 0.5$) , long-range interactions promote strong spin coherence in the ferromagnetic phase ($h_i < h_c$) for short chains, followed by a sharp decline for larger $N$, resulting in a negative ferromagnetic exponent ($a = -0.1035$) in Fig.~\ref{Fig14}-a . This is attributed to a collapse of spin alignment due to heightened susceptibility to thermal and quantum perturbations. Longer chains exhibit modest growth, as long-range interactions enhance fluctuations across a greater number of spins, manifesting more prominently in larger systems with increased resilience post-phase transition. For a high $\alpha = 1.2$ ($T_h = 5.0$, $T_c = 1.11$)  Fig.~\ref{Fig14}-b , short-range interactions yield small, consistent exponents ($a \approx 0.2355$) across all phases and all $N$, producing stable growth due to the rapid decay of perturbative long-range interactions. This restricts spin coherence, mitigating abrupt losses at $h_c$ and favoring the ferromagnetic phase.A pronounced amplification of the critical peak becomes apparent when examining the dependence of performance on system size. As previously discussed, the paramagnetic peak in \( \Pi/N \) exhibits a standard linear growth behavior, i.e., \( \Pi \sim N \), particularly in the case of long-range interactions characterized by the parameter \( \alpha \). In contrast, the peak associated with the critical transition displays a markedly superlinear increase for short-range interactions, while it is suppressed in the presence of long-range, nonlocal interactions.Similarly, the refrigerator exhibits analogous behavior in Fig.~\ref{Fig14}.c-d , resulting in an increased linearity of $\Pi_R/N$ in the ferromagnetic phase for $\alpha = 1.2$, with a slope coefficient rising from 0.47 to 0.94(in Fig.~\ref{Fig14}-d) . This indicates a size-dependent treatment across both phases, with a less pronounced drop compared to the engine mode.In systems with long-range interactions, paramagnetic peaks are enhanced and gradually decrease as the system size increases, eventually saturating to a finite value. This behavior contrasts with that observed for short-range interactions, where the maximum values of the critical peaks follow a clearly superlinear power-law scaling, well described by:
\[
\Pi_{\text{crit}}^{(R)}/N \sim N^{\alpha}, \quad (\alpha > 0).
\]

In conclusion, the performance of quantum thermal machines, whether as engines or refrigerators, is intricately linked to the scaling behavior governed by $\alpha$. High values of $\alpha$ in engines suggest promising avenues for enhancing efficiency through system size, particularly near critical points where spin correlations are maximized. However, refrigerators, even with intermediate values $\alpha$, show limited improvement, pointing to the need for optimizing interaction dynamics and minimizing dissipation to achieve better performance. Understanding these scaling laws and their implications can guide the design and optimization of next-generation quantum thermal devices, tailored for specific operational needs.

\section{Conclusion}
In this study, we investigated the performance of a long-range Ising chain quantum Otto engine and refrigerator under ideal conditions, focusing on the effects of system size \(N\), the interaction range parameter \(\alpha\), and the temperature ratios \(T_c/T_h\). Our analysis demonstrates that thermal correlations significantly influence the operational efficiency of these quantum machines, particularly near the critical points of phase transitions. We examined the behavior of all types of correlations for a long chain (N=100). For the quantum heat engine, long-range correlations (small \(\alpha\)) enhance performance in the precritical region, while short-range correlations (larger \(\alpha\)) dominate in the post-critical region. Similarly, for the refrigerator, long-range correlations are also favored in the post-critical field, especially in terms of cooling capacity \(Q_c/N\) and efficiency \(\eta_R\).We then fixed \(\alpha\) and varied \(N\) to investigate the influence of chain size on performance. For the quantum refrigerator, our results indicate that the cooling efficiency is maximized for smaller system sizes before the phase transition. At the critical point, the behavior of the efficiency \(\eta_R\) strongly depends on the value of \(\alpha\). For long-range correlations, efficiency decreases for short chains but increases for long chains. In contrast, for short-range correlations, \(\eta_R\) increases, showing better performance for short chains, while long chains exhibit greater stability and reduced dissipation. For the quantum heat engine, we observed relatively stable performance with minimal efficiency loss when the thermal gap is close and transitions occur before the phase transition, where long-range and intermediate correlations show better performance, even as the system size increases. This stability is attributed to the absence of non-adiabatic effects and the slow, quasi-static nature of the transformation. Additionally, the yield \(\eta\) differs from that of the refrigerator, as it depends on the chain size for all values of \(\alpha\).Overall, this study highlights the importance of distinguishing between the critical phase and the paramagnetic phase in understanding the performance of quantum heat engines and refrigerators. The ferromagnetic phase, characterized by significant fluctuations and superior performance, contrasts with the paraagnetic phase, which occurs beyond the critical point and exhibits greater stability but inferior performance. These findings underscore the need for a deep understanding of both static and dynamic conditions to effectively optimize the operation of quantum thermal machines.

\appendix
\section{Long range chain Ising}
\label{appendixA}
As a prototypical example of a many-body working substance with, possibly, long-range interactions, we consider a generic model of spinless fermions hopping across the $N$ sites of a linear chain in the presence of pairing interaction, and with a chemical potential $h$. Assuming periodic boundary conditions, the system Hamiltonian reads
\begin{align}
H =& -\sum_{j=1}^{N} \sum_{r=1}^{N/2-1} \left[ t_r c^\dagger_{j+r} c_j + \Delta_r c^\dagger_{j+r} c^\dagger_j + \text{h.c.} \right] \notag\\&- h \sum_{j=1}^{N} \left[ 1 - 2 c^\dagger_j c_j \right], \label{eq:hamiltonian}
\end{align}
where $c^\dagger_j$ and $c_j$ are creation and annihilation operators for fermions at site $j$, while $t_r$ and $\Delta_r$ are the hopping and pairing amplitudes, respectively. We choose their dependence on the intersite distance $r$ according to the power laws
\begin{equation}
t_r = \frac{1}{N_{\alpha}} \frac{J}{r^{\alpha}}, \quad \Delta_r = \frac{1}{N_{\alpha}} \frac{J}{r^{\alpha}}, \label{eq:power_laws}
\end{equation}
with the hopping exponent $\alpha_1 > 0$, the pairing exponent $\alpha_2 > 0$, and $N_\alpha = \sum_{r=1}^{N/2} r^{-\alpha}$ the Kac scaling factor, which guarantees extensivity of the energy in the case $\alpha_i < 1$, with $i = 1, 2$. Hereafter, we set $J = \Delta = 1$ as the energy scale and work in units of $\hbar = k_B = 1$. In particular, in the short-range case with $\alpha \to \infty$, the relation becomes exact through the Jordan--Wigner mapping anad we descibes the ising transversed field model \cite{Mbeng2024}. 

Due to the translational invariant nature of the couplings, it is useful to write the Hamiltonian in terms of the momentum-space operators
\begin{equation}
\tilde{c}_k = e^{-i \frac{\pi}{4}} \frac{1}{\sqrt{N}} \sum_{j=1}^N e^{ikj} c_j, \label{eq:momentum_operators}
\end{equation}
where $k = 2\pi n / N$, with $n = -N/2 + 1, \dots, N/2$ (in the following we will drop the $\tilde{}$ on the $c_k$ unless it is ambiguous). Then we obtain
\begin{equation}
H = \sum_k \left[ (h - t_k) \left( c^\dagger_k c_k - c_{-k} c^\dagger_{-k} \right) + \Delta_k \left( c^\dagger_k c^\dagger_{-k} + c_{-k} c_k \right) \right], \label{eq:hamiltonian_momentum}
\end{equation}
where $t_k$ and $\Delta_k$ are the Fourier transforms of the hopping and pairing amplitudes, respectively, which in the thermodynamic limit may be written as
\begin{align}
t_k &= \frac{\Re \left[ \text{Li}_{\alpha_1} \left( e^{ik} \right) \right]}{\zeta(\alpha_1)}, \label{eq:tk} \\
\Delta_k &= \frac{\Im \left[ \text{Li}_{\alpha_2} \left( e^{ik} \right) \right]}{\zeta(\alpha_2)}, \label{eq:delta_k}
\end{align}
where $\text{Li}_\alpha(z)$ denotes the polylogarithm and $\zeta(\alpha)$ is the Riemann zeta function. We notice that the Hamiltonian in the Fourier space can be decomposed into the sum of single mode Hamiltonians, introducing $\Psi_k = (c_k, c^\dagger_{-k})^T$
\begin{equation}
H = \sum_k \Psi^\dagger_k H_k \Psi_k, \label{eq:hamiltonian_decomposed}
\end{equation}
\begin{equation}
H_k = (h - t_k) \sigma^z_k + \Delta_k \sigma^x_k, \label{eq:mode_hamiltonian}
\end{equation}
where $\sigma^{(a)}_k$, $a = x, y, z$ are the Pauli sigma operators. Let us notice how the $k$th term of the Hamiltonian acts on a different sector of the total Hilbert space, namely the two-dimensional subspace spanned by the states $\ket{0_k, 0_{-k}}$, $\ket{1_k, 1_{-k}} = c^\dagger_k c^\dagger_{-k} \ket{0_k, 0_{-k}}$. Then, the Hamiltonian is diagonalized via a Bogoliubov transformation, in terms of the fermionic quasiparticle operators $\gamma_k = u_k c_k + v^*_{-k} c^\dagger_{-k}$, with Bogoliubov coefficients
\begin{equation}
u_k = \cos \frac{\theta_k}{2}, \quad v_k = \sin \frac{\theta_k}{2}, \label{eq:bogoliubov_coeffs}
\end{equation}
where $\theta_k = \arctan \left[ \Delta_k / (h - t_k) \right]$, to obtain
\begin{equation}
H = \sum_k \omega_k(h) \left( \gamma^\dagger_k \gamma_k - \frac{1}{2} \right), \label{eq:diagonalized_hamiltonian}
\end{equation}
with the spectrum
\begin{equation}
\omega_k(h) = 2 \sqrt{(h - t_k)^2 + \Delta_k^2}. \label{eq:spectrum}
\end{equation}
\section{The Nambu formalism}
\label{appendixB}

As we have seen, in the ordered case the Hamiltonian can be diagonalized by a Fourier transformation, reducing the problem to a collection of $2 \times 2$ ``pseudo-spin-1/2'' problems, followed by a Bogoliubov transformation, . In the disordered case, we can proceed similarly, but we cannot reduce ourselves to $2 \times 2$ problems in a simple way.
By using the Nambu formalism, we define a column vector $\bm{\Psi}$ and its Hermitian conjugate row vector $\bm{\Psi}^\dagger$, each of length $2L$, by
\begin{equation}
\bm{\Psi} = \begin{pmatrix}
\hat{c}_1 \\
\vdots \\
\hat{c}_L \\
\hat{c}^\dagger_1 \\
\vdots \\
\hat{c}^\dagger_L
\end{pmatrix} = \begin{pmatrix}
\hat{\mathbf{c}} \\
\hat{\mathbf{c}}^\dagger
\end{pmatrix},  \label{eqnambu_vector}
\end{equation}

$\bm{\Psi}^\dagger = \left( \hat{c}^\dagger_1, \dots, \hat{c}^\dagger_L, \hat{c}_1, \dots, \hat{c}_L \right) = \left( \hat{\mathbf{c}}^\dagger, \hat{\mathbf{c}} \right),$ or $\Psi_j = \hat{c}_j$, $\Psi_{j+L} = \hat{c}^\dagger_j$ and $\Psi^\dagger_j = \hat{c}^\dagger_j$, $\Psi^\dagger_{j+L} = \hat{c}_j$ for $j \leq L$.

\textbf{Warning:} Notice that the $\bm{\Psi}$ satisfy quite standard fermionic anti-commutation relations
\begin{equation}
\{\Psi_j, \Psi^\dagger_{j'}\} = \delta_{j,j'}, \label{eq:anticommutation}
\end{equation}
for $j, j' = 1, \dots, 2L$, except that $\{\Psi_j, \Psi_{j+L}\} = 1$ for all $j \leq L$, which brings about certain factors 2 in the Heisenberg's equations of motion (see later).

It is useful, for later purposes, to introduce the $2L \times 2L$ swap matrix $S$:
\begin{equation}
S = \begin{pmatrix}
0_{L \times L} & 1_{L \times L} \\
1_{L \times L} & 0_{L \times L}
\end{pmatrix}, \label{eq:swap_matrix}
\end{equation}
in terms of which $\bm{\Psi}^\dagger = (S \bm{\Psi})^T$.

Consider now a general fermionic quadratic form
\begin{equation}
\hat{H} = \sum_{j,j'} \left( A_{j'j} \hat{c}^\dagger_{j'} \hat{c}_j + A^*_{j'j} \hat{c}^\dagger_j \hat{c}_{j'} \right) + \sum_{j,j'} \left( B_{j'j} \hat{c}^\dagger_{j'} \hat{c}^\dagger_j + B^*_{j'j} \hat{c}_j \hat{c}_{j'} \right), \label{eq:quadratic_hamiltonian}
\end{equation}
where $A_{j'j} = A^*_{jj'}$, i.e., $A = A^\dagger$ is Hermitian, and $B_{jj'} = -B_{j'j}$, i.e., $B = -B^T$ is antisymmetric because $\hat{c}_j \hat{c}_{j'}$ is antisymmetric under exchange of the two operators, and any symmetric part of $B$ would not contribute. It is simple to verify that $\hat{H}$ can be expressed in terms of $\bm{\Psi}$, omitting an irrelevant constant term $\Tr A$, as:
\begin{equation}
\hat{H} = \bm{\Psi}^\dagger H \bm{\Psi} = \begin{pmatrix} \hat{\mathbf{c}}^\dagger, \hat{\mathbf{c}} \end{pmatrix} \begin{pmatrix} A & B \\ -B^* & -A^* \end{pmatrix} \begin{pmatrix} \hat{\mathbf{c}} \\ \hat{\mathbf{c}}^\dagger \end{pmatrix}. \label{eq:nambu_hamiltonian}
\end{equation}
There is an intrinsic particle-hole symmetry in a fermionic Hamiltonian having this form. This symmetry, further discussed in Sec.~6.1, is connected with the fact that the Hermitian $2L \times 2L$ matrix $H$ satisfies:
\begin{equation}
H S = -S H^*. \label{eq:particle_hole_symmetry}
\end{equation}

In the XY-Ising case, all couplings are real and we have two different fermionic Hamiltonians, one for each parity sector $p = 0, 1$, which we report here for convenience, using...

Let us consider the eigenvalue problem for a general Hermitian $2L \times 2L$ matrix $H$
\begin{equation}
H \begin{pmatrix} u_\mu \\ v_\mu \end{pmatrix} = \begin{pmatrix} A & B \\ -B^* & -A^* \end{pmatrix} \begin{pmatrix} u_\mu \\ v_\mu \end{pmatrix} = \epsilon_\mu \begin{pmatrix} u_\mu \\ v_\mu \end{pmatrix}, \label{eq:eigenvalue_problem}
\end{equation}
where $u, v$ are $L$-dimensional column vectors, composing the $2L$-dimensional column vector $\begin{pmatrix} u_\mu \\ v_\mu \end{pmatrix}$, and the $\mu$ index refers to $\mu$-th eigenvector. By explicitly writing the previous system, we find the so-called Bogoliubov-de Gennes equations:
\begin{equation}
\begin{cases}
A u_\mu + B v_\mu = \epsilon_\mu u_\mu, \\
-B^* u_\mu - A^* v_\mu = \epsilon_\mu v_\mu.
\end{cases} \label{eq:bogoliubov_de_gennes}
\end{equation}
It is easy to verify that if $(u_\mu, v_\mu)^T$ is eigenvector with eigenvalue $\epsilon_\mu$, then $(v^*_\mu, u^*_\mu)^T$ is an eigenvector with eigenvalue $-\epsilon_\mu$.
We can organize the eigenvectors in a unitary (orthogonal, if the solutions are real) $2L \times 2L$ matrix
\begin{equation}
U = \begin{pmatrix}
u_1 & \cdots & u_L & v^*_1 & \cdots & v^*_L \\
v_1 & \cdots & v_L & u^*_1 & \cdots & u^*_L
\end{pmatrix} = \begin{pmatrix}
U & V^* \\
V & U^*
\end{pmatrix}, \label{eq:eigenvector_matrix}
\end{equation}
$U$ and $V$ being $L \times L$ matrices (real, as we can choose to be, in the Ising case) with the $u_j$ and $v_j$ as columns. As a consequence:
\begin{align}
U^\dagger H U =& \begin{pmatrix}
\epsilon_1 & 0 & \cdots & 0 & 0 & 0 & \cdots & 0 \\
0 & \epsilon_2 & \cdots & 0 & 0 & 0 & \cdots & 0 \\
\vdots & \vdots & \ddots & \vdots & \vdots & \vdots & \ddots & \vdots \\
0 & 0 & \cdots & \epsilon_L & 0 & 0 & \cdots & 0 \\
0 & 0 & \cdots & 0 & -\epsilon_1 & 0 & \cdots & 0 \\
0 & 0 & \cdots & 0 & 0 & -\epsilon_2 & \cdots & 0 \\
\vdots & \vdots & \ddots & \vdots & \vdots & \vdots & \ddots & \vdots \\
0 & 0 & \cdots & 0 & 0 & 0 & \cdots & -\epsilon_L
\end{pmatrix}\notag\\& \equiv \text{diag}(\epsilon_\mu, -\epsilon_\mu) = E_{\text{diag}}. \label{eq:diagonalized_h}
\end{align}
If we define the new Nambu fermion operators $\bm{\Phi}$ and $\bm{\Phi}^\dagger$ in such a way that
\begin{equation}
\bm{\Psi} = U \bm{\Phi}, \label{eq:nambu_transformation}
\end{equation}
we can write $\hat{H}$ in diagonal form
\begin{equation}
\hat{H} = \bm{\Psi}^\dagger H \bm{\Psi} = \bm{\Phi}^\dagger U^\dagger H U \bm{\Phi} = \bm{\Phi}^\dagger E_{\text{diag}} \bm{\Phi}. \label{eq:diagonal_hamiltonian}
\end{equation}
Similarly to $\bm{\Psi}$, we can define new fermion operators $\hat{\gamma}$ such that
\begin{equation}
\bm{\Phi} = \begin{pmatrix}
\hat{\gamma} \\
\hat{\gamma}^\dagger
\end{pmatrix} = U^\dagger \bm{\Psi} = \begin{pmatrix}
U^\dagger & V^\dagger \\
V^T & U^T
\end{pmatrix} \begin{pmatrix}
\hat{\mathbf{c}} \\
\hat{\mathbf{c}}^\dagger
\end{pmatrix}. \label{eq:gamma_operators}
\end{equation}
More explicitly, we can write:
\begin{equation}
\begin{cases}
\hat{\gamma}_\mu = \sum_{j=1}^L \left( U^*_{j\mu} \hat{c}_j + V^*_{j\mu} \hat{c}^\dagger_j \right), \\
\hat{\gamma}^\dagger_\mu = \sum_{j=1}^L \left( V_{j\mu} \hat{c}_j + U_{j\mu} \hat{c}^\dagger_j \right),
\end{cases} \label{eq:gamma_explicit}
\end{equation}
which can be easily inverted, remembering that $\bm{\Psi} = U \bm{\Phi}$, to express the $\hat{c}_j$ operators in terms of the $\hat{\gamma}_\mu$:
\begin{equation}
\begin{cases}
\hat{c}_j = \sum_\mu \left( U_{j\mu} \hat{\gamma}_\mu + V^*_{j\mu} \hat{\gamma}^\dagger_\mu \right), \\
\hat{c}^\dagger_j = \sum_\mu \left( V_{j\mu} \hat{\gamma}_\mu + U^*_{j\mu} \hat{\gamma}^\dagger_\mu \right).
\end{cases} \label{eq:c_in_gamma}
\end{equation}
Finally, $\hat{H}$ in terms of the $\hat{\gamma}$ operators reads, assuming we have taken $\epsilon_\mu > 0$:
\begin{equation}
\hat{H} = \sum_{\mu=1}^L \left( \epsilon_\mu \hat{\gamma}^\dagger_\mu \hat{\gamma}_\mu - \epsilon_\mu \hat{\gamma}_\mu \hat{\gamma}^\dagger_\mu \right) = \sum_{\mu=1}^L 2 \epsilon_\mu \left( \hat{\gamma}^\dagger_\mu \hat{\gamma}_\mu - \frac{1}{2} \right), \label{eq:final_hamiltonian}
\end{equation}

\section{Adiabatic Transformation}

The time integral can be evaluated (assuming \(t\) is large) as:
\begin{equation}
\int_0^t e^{\frac{i}{\hbar} \Delta_{mn} t'} \, dt' \approx \frac{\hbar}{\Delta_{mn}} \left[ e^{\frac{i}{\hbar} \Delta_{mn} t} - 1 \right] \approx \frac{2\hbar}{\Delta_{mn}} \sin\left( \frac{\Delta_{mn} t}{2\hbar} \right)
\end{equation}

The transition amplitude becomes:
\begin{equation}
a_{n \to m}(t) \approx -\frac{2i}{\Delta_{mn}} \langle m(\lambda) | \frac{dH(\lambda)}{d\lambda} | n(\lambda) \rangle \frac{d\lambda}{dt} \sin\left( \frac{\Delta_{mn} t}{2\hbar} \right)
\end{equation}

The transition probability \(P_{n \to m}\) is the square of the transition amplitude:
\begin{align}
P_{n \to m}(t) &= |a_{n \to m}(t)|^2 \approx \left( \frac{2}{\Delta_{mn}} \right)^2 \left| \langle m(\lambda) \middle| \frac{dH(\lambda)}{d\lambda} \middle| n(\lambda) \rangle \right|^2 \nonumber \\
&\quad \times \left( \frac{d\lambda}{dt} \right)^2 \sin^2\left( \frac{\Delta_{mn} t}{2\hbar} \right)
\end{align}

For small transitions, the oscillating term \(\sin^2\left( \frac{\Delta_{mn} t}{2\hbar} \right)\), which oscillates between 0 and 1, can be ignored. Thus, the transition probability simplifies to:
\begin{equation}
P_{n \to m} \propto \left( \frac{1}{\Delta_{mn}^2} \right) \left| \langle m(\lambda) | \frac{dH(\lambda)}{d\lambda} | n(\lambda) \rangle \right|^2 \left( \frac{d\lambda}{dt} \right)^2
\end{equation}

% Create a figure environment that spans both columns at the top of the page

We have demonstrated that the transition probability \(P_{n \to m}\) is proportional to:
\begin{equation}
P_{n \to m} \propto \left( \frac{1}{\Delta_{mn}^2} \right) \left| \langle m(\lambda) | \frac{dH(\lambda)}{d\lambda} | n(\lambda) \rangle \right|^2 \left( \frac{d\lambda}{dt} \right)^2
\end{equation}
This result forms the basis of the adiabatic condition, which states that for a process to be adiabatic, the rate of change \(\frac{d\lambda}{dt}\) must be much smaller than the square of the energy difference \(\Delta_{mn}^2\), thereby ensuring that the transition probability remains low.

\section*{ACKNOWLEDGMENTS}

A.H. acknowledges the financial support of the National Center for Scientific and Technical Research (CNRST) through the “PhD-Associate Scholarship-PASS” program.
\par

\textbf{Declaration of Competing Interest:} 
\par
The authors declare that they have no known competing financial interests or personal relationships that could have appeared to influence the work reported in this manuscript.
\par
\textbf{Data Availability:}
\par
No data were used for the research described in this article.

\end{document}